\documentclass[12pt]{article}

\newcommand{\vctr}[1]{\ensuremath{\mathbf{ #1 }}}
\newcommand{\pb}[2]{\ensuremath{\left\{ #1 , #2 \right\} }}
\newcommand{\dr}[1]{\ensuremath{\mathrm{d} #1\,}}
\newcommand{\mc}[1]{\ensuremath{\mathcal{#1}}}
\newcommand{\dbd}[2]{\ensuremath{\frac{\dr{#1}}{\dr{#2}}}}
\newcommand{\vbv}[2]{\ensuremath{\frac{\delta #1}{\delta #2}}}
\newcommand{\ket}[1]{\ensuremath{\left|  #1 \right\rangle}}
\newcommand{\bra}[1]{\ensuremath{\left\langle #1 \right|}}
\newcommand{\bk}[2]{\ensuremath{\left\langle #1 | #2 \right\rangle}}
\newcommand{\proj}[2]{\ensuremath{\ket{#1} \bra{#2}}}
\newcommand{\tpk}[2]{\ensuremath{\ket{#1}\!\otimes\!\ket{#2}}}
\newcommand{\matel}[3]{\ensuremath{\bra{#1} #2 \ket{#3}}}
\newcommand{\hilbert}[1]{\ensuremath{\mathcal{#1}}} 
\newcommand{\op}[1]{\ensuremath{\widehat{\textsf{\ensuremath{#1}}}}}
\newcommand{\opad}[1]{\ensuremath{\op{#1}^{\dagger}}}
\newcommand{\id}{\op{\mathsf{1}}}

\newcommand{\mtr}[4]{\ensuremath{\left( \begin{array}{cc} #1 & #2 \\
#3 & #4 \end{array} \right) }}
\newcommand{\comm}[2]{\ensuremath{\left[ #1 , #2 \right]}} 
\newcommand{\tr}{\textsf{Tr}}
\newcommand{\nrm}{\frac{1} {\sqrt{2} } }

\newcommand{\be}{\begin{equation}}
\newcommand{\ee}{\end{equation}}

\newcommand{\ie}{\mbox{i.\,e.\,\ }}
\newcommand{\iec}{\mbox{i.\,e.\,}}

\newcommand{\egc}{\mbox{e.\,g.\,}}
\newcommand{\etc}{etc.\,\ }

\usepackage{chicago}

\begin{document}
\title{In defence of naivet\'{e}: The conceptual status of Lagrangian
quantum field theory}
\author{David Wallace}
\date{December 23, 2001}
\maketitle
\begin{abstract}
I analyse the conceptual and mathematical foundations of 
Lagrangian quantum field theory (that is, the `naive' quantum field theory
used in mainstream physics, as opposed to algebraic quantum field theory).  
The objective is to see whether Lagrangian quantum field theory has a sufficiently
firm conceptual and mathematical basis to be a legitimate object of foundational study,
or whether it is too ill-defined.  The analysis covers renormalisation and infinities,
inequivalent representations, and the concept of localised states; the conclusion is
that Lagrangian QFT (at least as described here) is a perfectly respectable physical theory,
albeit somewhat different in certain respects from most of those studied in foundational work.
\end{abstract}

\section{Introduction}

From its beginning, quantum field theory (QFT) has been plagued with
mathematical difficulties.  Any attempt to apply the theory to
interacting systems led to the appearance of infinities, and although
methods were found\footnote{By Dyson, Feynmann, Schwinger and Tomonaga, amongst others; see 
Cao (\citeyearNP{cao}, pp.\, 185--209) for a historical discussion.} to remove those infinities from the calculated 
cross-sections by cutting off certain integrals at finite (though very
high) energies,
these methods had a sufficiently \textit{ad hoc} feel
that the theory was not felt to be on remotely satisfactory conceptual
ground.

From the 1950s onwards, there has been a major program to resolve this
problem by reformulating QFT on an axiomatic basis: that is, starting
from what seem to be physically necessary --- and mathematically precise ---
principles which any QFT would have to satisfy, and then finding QFTs
which actually satisfy them.  This program is now generally referred to
as \emph{algebraic quantum field theory}, or AQFT; see \citeN{haag}, and
references therein, for extensive discussion of it.  

The major problem with AQFT is that very few concrete theories
have been found which satisfy the AQFT axioms.  To be precise, the only
known theories which do satisfy the axioms are interaction-free: no
examples are known of AQFT-compatible interacting field theories, and in
particular the standard model cannot at present be made AQFT-compatible.

Nonetheless, the great majority of foundational discussions of QFT are carried out in an AQFT framework,
for obvious reasons: the precise axiomatic nature of AQFT makes it ideal
for foundational study, since we can specify with mathematical precision
exactly what the entity is that we are studying.  

Meanwhile, the mainstream physics community has continued to work with
what we might call (for want of a better name) ``Lagrangian'' QFT, the
original QFT developed in the 1930s and whose infinities were tamed (at least pragmatically speaking) in
the 1950s.  In Lagrangian QFT, an entirely different program exists 
for understanding those infinities, in terms of the
renormalisation group; this program, due primarily to Wilson, does not
eliminate the finite cutoffs introduced to deal with the infinities, but rather
attempts to `legitimize' them.   It is in the framework of Lagrangian QFT
that the Standard Model is formulated.

This paper is an investigation of  whether Lagrangian QFT is
sufficiently well-defined conceptually and mathematically that it too
can be usefully subjected to foundational analysis.  The reasons for
making such an investigation are threefold.  Firstly, the problem with
restricting our foundational studies to AQFT is that --- pending the
discovery of a realistic interacting AQFT --- we have only limited
reason to trust that our results apply to the actual world, which
appears to be described rather well by the Standard Model.  

One response to this criticism is to say that Lagrangian QFT is so
mathematically ill-defined that we cannot regard it as a proper theory
at all, so we cannot trust its results either!  Our second reason for
the investigation is then to see whether this criticism is justified.

The third motivation is somewhat more philosophical.  Wilson's
explanation of the renormalisation procedure relies upon the failure of
the QFT to which it is applied at very short distances.  It is then
 intriguing to ask how to put on a firm conceptual footing a
theory which relies for its mathematical consistency on its own eventual
failure.

The structure of the paper is as follows.  Section \ref{whatis} reviews
the formal definition of a Lagrangian QFT.  This definition is very
formal and mathematically totally ill-defined; sections \ref{renorm} and
\ref{represent} address this problem, dealing respectively with
renormalisation and with the existence of inequivalent representations
of the observable algebra.  Section \ref{local} considers the problem of
how to define local objects in QFT.  This is also a problem in AQFT and
much of this section's argument applies in that domain; however, the
framework of sections \ref{whatis}--\ref{represent} offers some interesting
insights.  Section \ref{conclusion} is the conclusion.

Three comments should be made before the main part of the paper:

\begin{enumerate}
\item This paper is not intended as an attack on the AQFT program.  The
program to construct an AQFT-compatible interacting field theory is
well-motivated and, if successful, would be of enormous import; the
foundational results which have emerged from AQFT have been of
considerable importance in understanding quantum field theory and in
general they apply also to Lagrangian QFTs.  This paper should be read
as complementary to, rather than in competition with, these results.
\item There is a strong tradition in foundational studies of QFT to
treat the theory from an operational viewpoint, formalised by regarding
the local algebras of AQFT as describing operations which can be carried
out by an observer localised in a given region.  This paper does not
follow that viewpoint: it treats QFT instead as a closed theory (a
`universal theory' in Deutsch's \citeyear{deutsch85} sense), and regards
measurement devices, observers \etc as just subsystems of the QFT state.
This difference of approach has only limited significance for most of
the paper but is important in section \ref{loccon}.
\item Although much of the discussion below applies equally to bosonic
and fermionic field theories, the specific examples discussed are 
exclusively bosonic. 
\end{enumerate}

\section{What is a Lagrangian QFT?}\label{whatis}

For better or for worse, most canonical quantum field theories are found
by starting with a classical field theory and then `quantizing' it.  To
be sure, there is something intellectually unsatisfactory about this:
given that quantum theory is the more fundamental theory, we would
prefer to work in the other direction, that is, to recover
\emph{classical} field theories from quantum starting points (see
\citeN{deutschqft} for a development of this criticism).  Nonetheless,
the classical starting point has proven a powerful method for finding
QFTs, and we adopt it here.

\subsection{Classical field theories}

A classical relativistic field theory can be considered as consisting of a relativistic 
spacetime \mc{M}
(such as Minkowski spacetime),  a set of fields (that is, maps from \mc{M} 
to some other space, such as the real or complex
numbers\footnote{In more advanced treatments, we might take the fields
to be sections of some fibre bundle over \mc{M}; this
technicality does not significantly affect the arguments of the paper,
and will be ignored.}), and a Lagrangian density: a real-valued function
\mc{L} on \mc{M} whose value at a given point $x$ depends only on the
fields and their first partial derivatives, evaluated at $x$.  For
a given region \mc{D} of \mc{M}, we define the action $S_\mc{D}$ as
the integral of \mc{L} over \mc{D}; $S_\mc{D}$ is thus a functional of
the fields, and we define the dynamically allowed field configurations
within \mc{D} to be those for which $S_\mc{D}$ is extremal under
variations of the fields which vanish at the boundary of \mc{D}.

To go further, it is necessary to take \mc{M} to be globally
hyperbolic, and then to fix a foliation $\mc{M}=\Sigma \times \mathrm{R}.$  Let
$\Sigma_1, \Sigma_2$ be any two spatial slices in the foliation, with
associated time coordinates $t_1, t_2$ respectively; let the spacetime
region between the two slices be $\mc{D}_{12}$.
If
the fields are required to fall off rapidly enough at spatial infinity,
then we can define the action $S_{\mc{D}_{12}}$:
\be S_{\mc{D}_{12}}= \int_{\mc{D}_{12}}\!\!\!\!\! \dr{^4 \mu} \mc{L}
\;= \int_{t_1}^{t_2}\dr{t} \int_\Sigma \dr{^3\mu}\!(t) \mc{L}\ee
where $\dr{^4 \mu}$ is the four-dimensional volume element on \mc{M}
and $\dr{^3 \mu}\!(t)$ is the induced three-dimensional volume element
on $\Sigma$.  As the notation indicates, the three-volume element is in
general time-dependent; however, when \mc{M} has a time-translation
symmetry then we can always choose a foliation such that this time-
dependence vanishes.

We can now define the Lagrangian $L(t)$ of the theory by
\be L(t) = \int_\Sigma \dr{^3\mu}\!(t) \mc{L}. \ee  $L$ is now a
functional of the fields on $\Sigma$ and their time-derivatives, and the
action $S_{\mc{D}_{12}}$ is given simply by
\be S_{\mc{D}_{12}} = \int_{t_1}^{t_2}\dr{t} L(t),\ee so the field theory is
now in Lagrangian form (albeit infinite-dimensional).  

The final step before quantisation is to transform from Lagrangian to
Hamiltonian form, which we do by a straightforward infinite-dimensional
generalisation of the Legendre transform in classical mechanics: to each
field $\phi_i$ we associate a conjugate momentum field $\pi_i$ by
\be \label{lt1}\pi_i(\vctr{x}) = \vbv{L}{\dot{\phi}_i(\vctr{x})},\ee
and we define the Hamiltonian $H$ by
\be H[\phi,\pi] = \sum_i\int_\Sigma \pi_i(\vctr{x}) \dot{\phi}_i(\vctr{x})
- L[\phi,\dot{\phi}],\ee
where $\dot{\phi}_i$ is defined implicitly in terms of the $\pi_i$, via
(\ref{lt1}).
Points in the phase space \mc{P} of the field are then given by
specifying sets of functions $(\phi_i,\pi_i)$ (for all $i$); it is easy to check that 
the dynamics are given by Hamilton's equations,
as in the finite-dimensional case:
\be \dot{\phi}_i(\vctr{x}) = \vbv{H}{\pi_i(\vctr{x})};\ee
\be \dot{\pi}_i(\vctr{x}) = - \vbv{H}{\phi_i(\vctr{x})}.\ee

The Poisson bracket on \mc{P} is given by
\be \pb{A[\phi,\pi]}{B[\phi,\pi]} = \sum_i
\int_\Sigma\dr{^3\vctr{x}}\left( \vbv{A}{\phi_i(\vctr{x})}\vbv{B}{\pi_i(\vctr{x})} -
\vbv{A}{\pi_i(\vctr{x})}\vbv{B}{\phi_i(\vctr{x})}\right),\ee
and $\phi_i$ and $\pi_i$ obey natural infinite-dimensional analogues of the
canonical relations:
$\pb{\phi_i(\vctr{x})}{\phi_j(\vctr{y})}=\pb{\pi_i(\vctr{x})}{\pi_j(\vctr{y})}=0$
and $\pb{\phi_i(\vctr{x})}{\pi_j(\vctr{y})}=\delta_{i,j}\delta(\vctr{x}-\vctr{y}).$

Before going on to quantise this theory, we make two observations:
\begin{enumerate}
\item Despite the somewhat casual approach of this section, classical
field theories have perfectly well-defined Lagrangian and Hamiltonian
forms, and the above analysis can be carried out in a fully rigorous
way; see chapter 7 of \citeN{woodhouse}, chapter 3 of \citeN{marsden} 
and references therein for such an analysis.
In this more rigorous approach it is necessary to replace the
point functionals $\phi_i(\vctr{x})$ and $\pi_i(\vctr{x})$ with smoother
functionals on phase space, and to eschew use of the delta-function
Poisson bracket between such point functionals; such objects have a
status not dissimilar to that of position eigenstates in 
non-relativistic quantum mechanics, in that they cannot easily be rigorously defined
but (if used with a little caution) are of great conceptual and calculational convenience.
\item The almost immediate introduction of a foliation of \mc{M} may
cause some readers to worry about violation of relativistic covariance.
However, recall that each point in \mc{P} defines a unique
trajectory through \mc{P}, so that we may set up an isomorphism between
phase-space and the space of all solutions to the field equations.  The
latter space \emph{is} covariant (in the sense that it can be defined without
use of any preferred foliation) and if desired all talk of `phase space' may
be reinterpreted as talk of solution space.
\end{enumerate}
Both of these points are developed further in \citeN{waldqft} (from an AQFT perspective, and in the
somewhat restricted context of free fields).

For convenience, in the remainder of this paper we assume that we are
dealing with a single real field, and so drop the subscripts on
$\phi$ and $\pi$; this has no significant consequences.

\subsection{Quantization: observables}\label{observables}

In an ideal world, quantization would work as follows: it would yield a
Hilbert space \mc{H}, together with a map $Q$ from the functions on \mc{P}
(which can be thought of as the classical observables) into the space
of self-adjoint operators on \mc{H}.  $Q$ would have the property that,
if $\{,\}$ is the classical Poisson bracket, then $Q(\{A,B\} = i
(Q(A)Q(B) -Q(B)Q(A))$.

This is not an ideal world: quantization as defined above is provably
impossible (see, \egc, \citeN{abrahammarsden} for a discussion).  What is
possible is to find such a $Q$ for a much more restricted class of
observables: in nonrelativistic particle mechanics, for instance, we
choose spatial position and the  momentum conjugate to it.  There is no known
algorithm for choosing this restricted class, and no real reason to expect
one to exist (c.\,f.\, the criticisms of quantization in
\citeNP{deutschqft}); two different choices will sometimes lead to
empirically inequivalent quantum theories, in which case experiment is
the only way to determine which is correct.\footnote{\citeN{carlip} points out (p.\,97) that
`old quantum theory' can be (somewhat anachronistically) understood as wrongly selecting action-angle
variables as the preferred observables.}  In general, however,
Nature is fairly kind to us, and making the most obvious choice of
observables tends to work.  QFT is no exception: choosing $\phi(\vctr{x})$ and
$\pi(\vctr{x})$ as preferred observables leads to the empirically
correct theory.

The question of which observables count as `fundamental' is also
relevant for the interpretation of the quantum theory which we hope to
produce.  For any fixed dimension, all Hilbert spaces are isomorphic, 
so it is through the observables that we are able to give physical
meaning to a theory.  In practice, this usually involves a connection
between certain observables and the spacetime of the theory: in NRQM,
for instance, it is the designation of the momentum operators as
generators of translations which allows us to identify certain states as
eigenstates of position.\footnote{The argument, adapted from
\citeN{newtonwigner}, is as follows (in one dimension): if a state \ket{\phi} is localised at some
point, then any translation $\exp (i \lambda \op{P})$ applied to it will
leave it localised at a different point; thus $\exp (i \lambda
\op{P})\ket{\phi}$ must be orthogonal to \ket{\phi}; hence the position
eigenstates are precisely those satisfying $\matel{\phi}{\exp (i \lambda
\op{P})}{\phi}$ for all $\lambda$.  If $\{\ket{k}\}$ are the (improper)
eigenstates of \op{P}, and if we define 
$\ket{x} = \int\dr{k}\exp(-i k x)\ket{k}$, then $\bk{x}{y}=\delta(x-
y)$, and $\exp (i \lambda \op{P})\ket{x}=\ket{x-\lambda}$; hence, $\matel{x}{\exp (i \lambda
\op{P})}{x}= \bk{x}{x-\lambda}$, and so the
\ket{x} states are exactly localised.}  The obvious strategy in QFT is
to use the spatiotemporal dependence of the observables to give the
theory physical meaning: when quantised, $\phi(\vctr{x})$ and
$\pi(\vctr{x})$ will become operators which we will treat as localised
at $\vctr{x}$.  

This approach can be taken further.  First we shift from the
Schr\"{o}dinger to the Heisenberg picture, so that the field operators
are functions of spacetime points $x$ rather than just spatial points
$\vctr{x}$.  Then, given any (bounded, open) subset
\mc{O} of \mc{M}, we define the algebra $\mc{A}(\mc{O})$ as the algebra
of all operators which can be constructed from $\op{\phi}(x)$ and $\op{\pi}(x)$
whenever $x\in \mc{O}$ --- so $\mc{O}$ includes such operators as 
$\op{\pi}(x)^2 (x\in \mc{O})$,
$\int_\mc{O}f(x) \op{\phi}(x)$, etc.  This gives us (formally) what is
technically known as a \emph{net} of operator algebras: a map from bounded, open subsets of
a topological space into operator algebras, such that if $\mc{O}_1
\subseteq \mc{O}_2$, then $\mc{A}(\mc{O}_1)$ is a subalgebra of
$\mc{A}(\mc{O}_2)$.
It is a central claim of algebraic quantum field theory --- and one
which appears consistent with the way QFT is used in physics --- that a QFT
is entirely specified by the structure of this algebra net: in other
words, that all physical facts about the theory can be determined from
knowing the spacetime dependence of the observables, without any need to
know (for instance) which observable is $\op{\pi}(x)$ and which is
$\op{\phi}(x)$.  In particular, this allows the possibility that field
theories generated from different classical Lagrangians are actually the
same QFT; two such field theories are generally referred to as
\emph{Borchers equivalent}.

As such, to specify an algebraic quantum field theory, all that is
needed is a net of algebras, which is required to satisfy certain axioms
(basically relating to locality and causality; see \citeN{haag} for an
extended discussion of these axioms).  Any such AQFT implicitly
specifies a Borchers equivalence class of `concrete' QFTs each of which generates 
the same algebra net; all such concrete QFTs are Borchers equivalent to
one another, and they are usually regarded simply as different ways of
describing the AQFT,
analogous to different coordinate systems on a manifold.

The assumption that the net of algebras captures all physical
information about a theory gives a very elegant realisation of the idea
that quantum theories gain physical content via a specification of the
spatiotemporal properties of their observables.  We will refer to it as
the Net Assumption.

We have glossed one question, though: what does it mean to say that an operator is
localised somewhere?  In much of the AQFT literature an operationalist
answer is given to this question: operators represent physical
operations which can be performed, by the observer, on the QFT state,
and in particular $\mc{A}(\mc{O})$ is the algebra of all operations
which can be performed, and all measurements which can be made, by an 
observer who is external to the field system but localised within $\mc{O}$.
However --- as mentioned in the Introduction --- we wish to understand
a given QFT as self-contained (in which case observers, measurement devices and the like must
be built out of states of the QFT) so this approach is not available to us.
Instead, we take the expectation values of operators in a given spatial
region as giving us information about the degree of excitation of the
field in that region.  This view will be further developed in the next
section.

\subsection{Quantization: states}\label{wavefunctional}

We now address the practical task of actually finding the operator
representations of the field observables.  In NRQM we usually do this via 
wave-functions: we take our Hilbert space to be the space of complex functions $\psi(q)$ on the
configuration space of the classical theory, and quantize the classical
observables $q,p$ via 
\be  \op{q} \psi =q \psi(q); \,\,\,\op{p} \psi = -i \dbd{\psi}{q}.\ee
Formally (and only formally; but see sections \ref{renorm} and
\ref{represent}) we can easily generalise this to infinite-dimensional
systems: the configuration space of the system is the infinite-dimensional 
space of all field configurations on $\Sigma$, so the quantum states will be complex-valued
functionals on that space.  We will denote the Hilbert space of such
functionals by $\mc{H}_\Sigma$, and proceed to quantize the classical
observables  $\phi(\vctr{x})$ and $\pi(\vctr{x})$ as
\be (\op{\phi(\vctr{x})} \Psi)[\chi] = \chi(\vctr{x})\Psi[\chi];\ee
\be (\op{\pi(\vctr{x})} \Psi)[\chi] = - i
\vbv{\Psi}{\chi(\vctr{x})}[\chi],\ee 
where $\chi$ is any field configuration on $\Sigma$. It is easy to check that 
infinite-dimensional generalisations of the 
canonical commutation relations are satisfied:
\be \comm{\op{\phi(\vctr{x})}}{\op{\phi(\vctr{y})}} = 
\comm{\op{\pi(\vctr{x})}}{\op{\pi(\vctr{y})}}=0;\ee
\be \comm{\op{\phi(\vctr{x})}}{\op{\phi(\vctr{y})}} = i
\delta(\vctr{x}-\vctr{y}).\ee
(For simplicity we have considered only the case of one real field, but
the generalisation to multiple fields is obvious.)

Now the Hilbert space $\mc{H}_\Sigma$ has a natural tensor-product
structure.  Let $\Sigma_1, \ldots, \Sigma_n$ be disjoint subsets of $\Sigma$
whose union is $\Sigma$; then 
(formally) we  can define $\mc{H}_j$ as the space of complex functionals on functions
on $\Sigma_j$.  Any function $f$ on $\Sigma$ can be specified uniquely by giving those $n$ functions
which are the restrictions of $f$ to each $\Sigma_j$; hence
we have
\be \mc{H}_\Sigma = \otimes_{i=1}^n \mc{H}_{\Sigma_i}.\ee
If $\vctr{x}\in \Sigma_j$ then $\op{\phi(\vctr{x})}$ and
$\op{\pi(\vctr{x})}$ act trivially on $\mc{H}_{\Sigma_i}$ for $i \neq
j$.  Hence, the $\mc{H}_{\Sigma_i}$ can be thought of as representing those subsystems of
the field which are localised within $\Sigma_i$.  This viewpoint can be reversed:
we can take spatially localised subsystems of the field as our starting
point, in which case $\op{\phi(\vctr{x})}$ and
$\op{\pi(\vctr{x})}$ are localised in $\Sigma_j$ \emph{because} they act
trivially on the other $\mc{H}_{\Sigma_i}$.

The reader is likely to have noticed that none of the above would win
prizes for mathematical rigor!  Indeed, on the face of it the approach of this whole
section is appallingly badly defined.  It will be the purpose of
sections \ref{renorm} and \ref{represent} to explain how we can attain
peaceful coexistence with this ill-definedness.

\section{Infinities and renormalisation}\label{renorm}

We now have at least a formal definition of a quantum field theory; but
it is, clearly, profoundly unsatisfactory in many respects.  In this
section we consider the question of how, and to what extent, it is
possible to restore some degree of mathematical rigour to the theory.

\subsection{The problem of infinities}

Where to start in analysing the mathematical shortcomings of section
\ref{wavefunctional}?  We could begin by noting the various ambiguities
glossed over when considering functionals on an infinite-dimensional
space: what topology should be placed on the space of functionals?  How
smooth must a field configuration be to be admitted?, etc.  However,
there is a more crucial problem which applies to any attempt to make
wave-functional space a Hilbert space.  This is the problem of how to
define the inner product.  In the finite case, of course, we define the
inner product of two wave-functions by
\be (\psi_1,\psi_2) = \int_{\Re^n} \dr{^n q}\psi_1^*(q_1, \ldots , q_n)
\psi_2(q_1, \ldots , q_n);\ee generalising this to wave-functionals
gives a functional integral:
\be (\Psi_1, \Psi_2) = \int_\mc{S} \mc{D}\!\chi
\Psi^*_1[\chi]\Psi_2[\chi],\ee
where $\mc{S}$ is field configuration space and $\mc{D}\chi$ is a
`functional measure' on $\mc{S}$.  

Unfortunately, defining measures on infinite-dimensional spaces is
extremely hard.  Indeed, simple attempts to define $\mc{D}\chi$ tend to give infinity as
the result of the functional integral, which clearly isn't satisfactory (see, \egc, 
Binney \textit{et al} \citeyear[p.\,409]{binney}
for an example of how this occurs).   The integral can be made 
well-defined if we restrict ourselves to a finite-dimensional subspace
of \mc{S}, but as soon as we start to consider functions which can vary
on arbitrarily short length-scales, we lose the ability to define it.

Blithely ignoring this little problem, we can press on with the
development of our QFT, but we soon run into other infinities:
in particular, when we try to calculate the effects of interactions, 
we find that our calculations include terms which involve integrating
over arbitrarily short length-scales, and that some of these terms
are infinite when the integral is taken over such length-scales.  With
care we can avoid the infinities in free-field theories (such theories
can be exactly defined in AQFT) but no examples are known of realistic
QFTs in which these infinities do not have to be confronted.  

One highly principled attitude to the problem might be to say: very
well, none of our supposed `interacting QFTs' count as real theories, so
let us reject all of them and go looking for properly defined ones.
Needless to say, though, this is not the mainstream approach
in particle physics, where algorithms for extracting useful information
from QFT despite the infinities have been known for nearly fifty years.
To understand why these algorithms work --- and why they can be
understood not just as `algorithms', but as a valid resolution of the
problem of infinities --- we take a brief digression into 
condensed-matter physics.

\subsection{High-energy cutoffs and renormalisation}

Functional integrals are not restricted to relativistic quantum theory:
they occur throughout physics, and in particular in condensed-matter
physics.  There too they formally lead to infinities; however, there is
no question of a conceptual problem.  For the integrals are taken
on the assumption that (for instance) the density of matter is
continuously varying, and can vary on arbitrarily short length-scales.
But this continuum assumption can only be an approximation, for matter
is made of atoms, and any variations on a length-scale short relative to
the interatomic distance clearly lie beyond the scope of the
approximation.  So the functional integral must be cut off at some
short, but finite, length-scale.

It might be thought that this would lead to enormous computational
difficulties, for the functional integral is dominated by the 
short-distance variations and hence the precise details of the cutoff
procedure ought to be important.  Remarkably, this is not so: it can be
shown that if we restrict our attention to the behaviour of the system
on lengthscales which are very long relative to the cutoff, then: 
\begin{enumerate}
\item All but a finite number of possible interaction terms (the 
so-called `renormalisable' terms) have negligible effect on the system's
behaviour.
\item The only effect of the high-energy (\ie short-distance) degrees of
freedom on the system's long-distance behaviour is to modify
(`renormalise') the coefficients of the renormalisable interaction
terms.  As such, the only effect of getting the cutoff details wrong is
to change the effective values of the coefficients in the Hamiltonian.
If, instead of trying to calculate these effective values based on the true
values (which in any case is often impossible) we simply measure them,
the details of the cutoff procedure are completely irrelevant to the
long-distance behaviour of the system.
\end{enumerate}
These results were established in the 1970s, primarily by Kenneth Wilson
(see \citeN{wilson} and references therein); Binney \textit{et al}~\citeyear{binney} give an
exceptionally lucid exposition.

At least mathematically, this process can be applied equally well to
relativistic QFTs: instead of allowing our functional integrals to range
over all field configurations, we restrict them to only those
configurations whose short-distance variation is not too quick (the
conceptually simplest way, at least in Minkowski spacetime, is to exclude 
all configurations whose Fourier coefficients vanish above some fixed value of 
$|\vctr{k}|$, where $1/|\vctr{k}|$ is the intended cutoff length).  This yields a well-defined 
theory, and if we require the cutoff length to be far smaller than the
lengthscales at which we study the theory, then that cutoff length
affects the theory's predictions only through renormalisation of the
coefficients in the effective Hamiltonian.\footnote{See chapter 12 of \citeN{peskinschroeder}
for the technical details of this process.} Since the latter
coefficients are in any case only known empirically, this has no
practical consequences. 

What physical justification might there be for imposing a cutoff in
relativistic QFT?  Three possibilities are generally mentioned in the
literature (see, \egc, \citeN{cao} or Binney \textit{et al}~\citeyear{binney} for details):
\begin{itemize}
\item It is easy to construct field theories in which some of the
degrees of freedom are `frozen out', \ie become irrelevant, below some
given energy scale.  Below this scale, the field theory can be described
by an `effective field theory' (EFT), which does not explicitly include the
frozen-out degrees of freedom; however, those degrees of freedom do have
the effect of imposing an effective cutoff of the EFT.  This is believed to happen for
a number of actually studied low-energy theories, whose cutoff energies are experimentally
accessible; see Binney \textit{et al}~\citeyear[pp.\,372--373]{binney}, for a brief discussion.  Of course, when
a cutoff is generated this way, it relies on the existence of the `true'
field theory, which in turn can only be rigorously defined with a cutoff
of its own.  This implies either some infinite tower of EFTs, which
describe physics at successively higher energy scales, or else some
other cutoff mechanism which truncates the tower.
\item Possibly at sufficiently high energy scales (say, beyond the
`Grand Unification energy' \cite{peskinschroeder}, about $10^{16}$ GeV, above which the strong and
electroweak forces are conjectured to become unified) the entire 
field-theoretic description of physics may break down and be replaced by some
other theory (such as string theory).  This new theory would then impose
an effective cutoff, and might itself be free of infinities.
\item At still higher energy levels (and hence still shorter lengthscales), it is widely believed 
that the concept of spacetime as a continuum will itself break down, to
be replaced by some quantized version.  On dimensional grounds this is
expected to happen at around the Planck energy, or $10^{19}$ GeV; this is equivalent to
a breakdown on lengthscales of order $10^{-34}$ metres (the Planck length).
\end{itemize}
In practice, the second and third alternatives may not be all that
distinct: although string theory is usually formulated as a
perturbative theory on a flat background spacetime, it is generally
accepted that this background spacetime must eventually be eliminated
from the theory --- in which case, presumably, we would again expect
spacetime to be an effective concept emerging only on lengthscales
greater than the Planck length.

\subsection{The conceptual status of a cutoff QFT}\label{cutoffconcept}

The introduction of a finite cutoff, then, mathematically resolves the
infinities problem, and might turn out to be justifiable on a number of
physical grounds.  However, it is prima facie unfortunate from a
foundational perspective, for the following reason: in foundational
work, it is usual to start with a well-defined (and, preferably,
axiomatisable) theory, and then to investigate the implications of that
theory.  If QFTs are intrinsically approximate theories, which can be
trusted only at certain energy scales and whose behaviour is unknown at
other scales, it is hard to take this attitude to them.   Should we then
regard a QFT as a well-defined theory at all, or (as \citeN{buchholz} has
advocated) as 
\begin{quote}
an efficient algorithm for the theoretical treatment of certain specific 
problems in high-energy physics \cite[pp.\,1--2]{buchholz}
\end{quote}
\citeN[pp.\,350--352]{cao} identifies three different, commonly adopted, attitudes to the 
foundations of QFT:
\begin{enumerate}
\item[(1)] The current situation is genuinely unsatisfactory: we should
reject the cutoff theories as not mathematically well-defined, and
continue to look for nontrivial theories defined at all lengthscales.
\item[(2)] The picture of `an infinite tower of effective field theories'
mentioned in the last section is to be taken seriously.  As Cao
stresses, this would require 
\begin{quote}a drastic change of our conception of fundamental physics
itself, a change from aiming at a fundamental theory (as the foundation
of physics) to having effective theories valid at various energy scales.
\cite[p.\,351]{cao}\end{quote}
\item[(3)] QFTs as a whole are to be regarded only as approximate
descriptions of some as-yet-unknown deeper theory, which gives a
mathematically self-contained description of the short-distance physics.
\end{enumerate}
(1) is of course the viewpoint of the entire algebraic QFT program: that
program's ultimate long-term goal is to produce a mathematically
rigorous description of interacting QFTs (specifically, of the Standard
Model or of some successor to it).  Obviously, completion of that goal
would dramatically change the foundational status of quantum field
theory; it is only because the goal has not been achieved that it is
interesting to investigate alternatives to (1).  Conversely, anyone regarding either
(2) or (3) as fully satisfactory from a foundational viewpoint would
have reason to doubt whether the eventual success of the AQFT program
is possible.

(2) will not be discussed further here: adopting it would
require dramatic changes to many aspects of our attitude to the
foundations of physics, and a discussion of such changes lies far beyond
this paper.  See \citeN{castellani} and \citeN{hartmann} for further discussion.

The rest of this discussion will focus on (3), and its attendant
foundational problem: how can we give a clean conceptual description of
a theory which can be rigorously defined only as the low-energy limit of
another theory which we do not yet have?

To see the problem here, contrast this case with a more straightforward
one: the current status of classical mechanics (CM), given that we believe it
to be in some sense superseded by non-relativistic quantum mechanics
(NRQM).  What is our justification for continuing --- as in many
situations we do continue --- to use CM, given that we believe NRQM to
be the more valid description?  It is not enough to observe that NRQM
tends to CM as $\hbar \rightarrow 0$, for $\hbar$ does not tend to zero:
it is a constant.  Furthermore, simply considering successively larger systems (so that
the quantity ($\hbar/$typical action) tends to zero) is singular and 
contains various surprises (this matter is discussed in \citeNP{zurek98}).
A better approach is to identify what might be called
`classical domains' of NRQM: that is, domains of NRQM in which NRQM is
approximately isomorphic to CM.  (A system of large bodies decohered by
interactions with an environment would be an example of such a domain.)
Within such a domain, we could then tentatively apply CM, whilst
remaining aware that CM could fail to give accurate answers to some
questions, and that its predictions should only be treated as accurate
to within the accuracy of the `approximate isomorphism' between CM and
the restricted domain of QM.
(This approach to inter-theoretic reduction is discussed in much more
detail, and in the specific context of the relationship between CM and
NRQM, in \citeN{wallacestatmech}; it is closely related to Cao's 
structural approach to QFT \cite[chapter 12]{cao} and to the
`structural realist' program inaugurated by \citeN{worrall} and
discussed in, \egc, \cite{psillos,ladyman}.)

Approach (3) presumes the existence of some theory --- call it X ---
to which QFT is to be an approximation (in some domains).  There are,
however, a number of important differences between the X--QFT
relationship and the NRQM--CM one:
\begin{enumerate}
\item Most obviously, we know NRQM, whereas we have only the sketchiest
idea of what X will turn out to be.  This makes it difficult to
delineate the domains in which QFT is approximately isomorphic to X,
since we have to describe those domains in QFT-language rather than 
X-language.  In particular, we would like to say that these domains are
those in which the states do not vary on too-short lengthscales, but
this implicitly suggests that it even makes sense to talk about states
which do vary on such lengthscales -- which may not be possible in X:
indeed, X may not contain any elements which are even approximately
isomorphic to too-rapidly-varying states of QFT.  In solid-state
physics, by analogy, it isn't even meaningful to talk about density
fluctuations on lengthscales shorter than the interatomic separation.
\item When using CM, in situations where we are unsure about its
validity we can always cross-check with NRQM to check that its
predictions agree with CM.  Since we don't have X, we can't do this in
QFT: the only tests available for its validity in describing a given
phenomenon are empirical. 
\item CM is a mathematically well-defined and self-consistent theory, whereas QFT as we have
described it so far is not.
\end{enumerate}
It is useful to separate the first and second points from the third.
For \emph{if} we had a mathematically well-defined QFT, we would have a
perfectly good physical theory which, however, we would have good reason
to believe empirically inadequate for describing certain phenomena, even
though we have no better alternative.
This has been a common situation throughout physics: classical mechanics
in the early 20th century is an example of such a theory; general
relativity today is another.  In fact, until and unless we come up with
a theory which we are supremely confident is `final', this should be our
attitude to any well-tested physical theory: its predictive utility and
explanatory power require us to take seriously those entities which the
theory claims to exist (such as classical particles, or 
general-relativistic spacetimes), but to remember that 
\begin{itemize}
\item these entities may be patterns or structures in some more
fundamental ontology, rather than fundamental entities in their own
right;
\item our theory may describe the world not by virtue of being
`fundamental', but by virtue of being instantiated in the structure of a
deeper theory;
\item this instantiation may be only approximate, leading to domains in
which the theory fails.
\end{itemize}
(See \citeN{wallacestructure} for a discussion of the ontology of
patterns and structures in physics.)

So, if we had a mathematically well-defined QFT then we could happily
interpret it in the same way we now interpret CM.  Can we obtain such a
QFT?

The easiest way to do so is to choose a concrete implementation of the
short-distance cutoff.  A simple recipe for such an implementation is as
follows:
\begin{enumerate}
\item Choose a cutoff lengthscale $l$.
\item Choose a spacetime foliation.
\item On each leaf of the foliation, replace the continuum of field
operators $\op{\phi}(\vctr{x})$, $\op{\pi}(\vctr{x})$ with a discrete
grid of such operators (to be called grid observables) 
with the grid spacing being $\sim l$.
\item Discretise the Hamiltonian, replacing its integral over functions
of field operators with a discrete sum over operators on the grid.
\item Replace the (formal) Hilbert space $\mc{H}_\Sigma$ of the QFT
(defined in section \ref{wavefunctional} as the space of complex
functionals over all field configurations on $\Sigma$) with the space of
functionals over some subset of field configurations, chosen to vary
in some suitably chosen way between points of the grid.  (The details of how this is to be
done don't matter --- \ie have no consequences for the large-scale
behaviour of the model --- provided that specifying the value of a field
configuration at all grid points is enough to identify it uniquely).
\end{enumerate}

Such a theory is a close relative of the QM description of a crystal: in
the latter, we specify the position and momentum of each crystal atom;
in the former, the field strength and conjugate momentum at each
gridpoint.  (It is also very similar to the ``lattice gauge theories''
studied in lattice QCD.)  It is also perfectly well-defined
mathematically, and as such a valid entity to be viewed as approximately
isomorphic to some subtheory of X.

We are interested in the structure of states of such theories, when
studied at lengthscales $\gg l$.  To give a precise meaning to this,
recall that the physical meaning of a state is given by its expectation
value on elements of the algebra of observables.  We can define the
\emph{large-scale} observables of a discretised QFT as those which are
averages of grid observables over regions which are very large compared
to the grid spacing; then we can regard the large-scale structure of a
state as specified by its expectation value with respect to all 
large-scale observables.\footnote{By `grid observables' I mean the basic
field observables $\op{\phi}(x)$ and $\op{\pi}(x)$, not their pointwise
products --- so for instance $\int \dr{x}\dr{y}f(x)g(y)
\op{\phi}(x)\op{\phi}(y)$ is a large-scale observable (if $f$ and $g$
vary slowly compared to $l$) but $\int \dr{x}f(x)
\op{\phi}(x)^2$ is not, irrespective of how smoothly $f$ varies.  This
means, in particular, that the energy density is not a large-scale
observable.  In fact, the energy density of a discretised QFT is (a)
cutoff-dependent, and (b) extremely large: this is the origin of the
`cosmological constant problem'.}
The small-scale structure of a state is not to
be treated as physically significant, since it is defined at
lengthscales at which we expect the approximate isomorphism between X
and our discretised QFT to break down.  (Analogously, the structure of a
classical phase-space distribution can be fairly reliably viewed as
telling us about the actual world if it is studied on action scales large
compared to $\hbar$, but is probably meaningless on scales small
compared to $\hbar$).

Now, obviously there is a very large number of possible discretisations
of a field theory: one for each possible choice of grid, at least (and
no claim is being made here that the method above is the only way of
implementing the cutoff).  But renormalisation tames this profusion of
theories to some extent, for it implies that the large-scale structure
of two different discretisations can be made to be virtually identical
(and totally identical in the limit as the two theories are studied on
larger and larger length-scales) by finite rescaling of the field observables 
and finite adjustments in finitely many
parameters of the Hamiltonians of the theories. 

By analogy with section \ref{observables}, we can construct an algebra
net by associating to each region $\mc{O}$ which is large compared with
the cutoff length $l$, an algebra $\mc{A}(\mc{O})$ of large-scale
observables localised in $\mc{O}$.  The statement that two discretised
QFTs are structurally equivalent at large scales is then equivalent to
the statement that both generate the same large-scale nets --- a 
large-scale variant of the Net Assumption of section \ref{observables}.

In fact, we can mimic the algebraic description of section
\ref{observables} further, by defining our cut-off QFTs directly in
terms of their algebraic structures: we define a \emph{scale-l algebraic QFT}
as a map from regions of \mc{M} which are large compared with $l$, to
operator algebras, such that the axioms of AQFT are approximately
satisfied on scales large compared with $l$.  (Obviously this is not
intended to be precise.)  Two `concrete' QFTs, such as two versions of the 
discretised QFT above, are then \emph{scale-l equivalent} if they
generate the same scale-l AQFT; hence scale-$l$ equivalence is a
generalisation of Borchers equivalence.  

The results of renormalisation theory can now be simply stated: two
discretisations of the same renormalisable QFT can be made scale-$l$
equivalent just by finite renormalisations of the coefficients in one of
them.  So, for $\phi^4$ QFT (for instance) there is a two-parameter
family of scale-$l$ AQFTs, parametrised by the renormalised mass and
coupling constant.  

Now if we know that a given scale-$l$ AQFT is empirically accurate, 
we can find any number of concrete QFTs which are members of its 
scale-$l$ equivalence class.  We can choose any such member to calculate
with, and make our choice based upon calculational convenience, but we need not regard
any of them as candidates for the fundamental theory of nature.
Rather, they are to be viewed as candidate
theories to be approximately isomorphic to a subtheory of X.  

From this perspective, the profusion of discretisations is no
embarassment.  For suppose $QFT_1$ and $QFT_2$ are members of the same scale-$l$ equivalence class,
and suppose the scale-$l$ AQFT corresponding to that equivalence class accurately describes our
observations (on scales large compared with $l$).  Then it follows that   
$QFT_1$ and $QFT_2$ are both approximately isomorphic to a subtheory of
X\footnote{Note that $QFT_1$ and $QFT_2$ must be approximately
isomorphic to the \emph{same} subtheory of X, since they are intended to
describe the same physical domains as one another.} with that
approximate isomorphism breaking down around the cutoff lengthscale.
But if so, then there can be no fact of the matter as to which of $QFT_1$ and
$QFT_2$ is `correct': \emph{both} are approximately instantiated by X.
If X is itself a QFT (presumably some interacting QFT which exactly
satisfies the axioms of AQFT) then X itself will be inside the scale-$l$
equivalence class, but if X is something else --- a string theory, or
some theory in which spacetime is quantized --- then no elements of the
equivalence class are fundamental: all are useful purely because of
their structural resemblance to a subtheory of X.

\subsection{Covariance}\label{covariance}

One of the least aesthetic features of discretised QFTs is their 
non-covariance, which manifests itself in (at least) two ways.  Firstly,
a lattice is inevitably not preserved under Poincare transformations
(indeed, it is not even preserved under translations).  Secondly, the
commutators of field operators which are spacelike separated are in
general not zero (although they drop off extremely rapidly if the field
operators are separated by distances large compared with the cutoff
distance).  This second effect is because of the discretisation of the
Hamiltonian, in which spatial derivatives like $\nabla \phi(\vctr{x})$ are
replaced with terms like $(\phi(\vctr{x})-\phi(\vctr{x}+\vctr{l}))/l$,
where $l$ is the cutoff length and $\vctr{l}$ is a vector of length $l$.
As a consequence, influences can propagate instantaneously from $\vctr{x}$ to
$\vctr{x}+\vctr{l}$, although the effect is miniscule for widely
separated points.

This is not to say that the approach advocated in section \ref{cutoffconcept}
implies that the actual world has faster-than-light signalling or a 
preferred reference frame.  Recall that in this approach, our
discretised QFT is taken to describe the actual world only insofar as it
is isomorphic to a subtheory of the deeper ``theory X'', and we expect
that this isomorphism will hold only for expectation values of field
observables averaged over regions large compared with the cutoff length.
But for these expectation values, we find that
violations of covariance are very small, tending to zero as the
regions over which the observables are averaged increase in size.
In other words, non-covariance is essentially a small-scale property of
discretised QFTs, and such properties are precisely those for which, it 
was argued, we should not regard QFT as telling us about the actual world.

Notwithstanding this, the situation is still rather unsatisfactory, at least if
we view covariance as a fundamental property of physics rather than some effective
limit of a non-covariant theory.  There are a number of ways of coming
to terms with this:
\begin{enumerate}
\item We might manage to find a covariant way of defining discretised
QFTs.  Obviously this would remove the covariance problem; however, it
seems rather unlikely since cutoffs are about short lengthscales, and
length is not a relativistic invariant.  (Of course, if we look for a
covariant QFT defined without a cutoff, we are moving away from the
approach described here, and back towards the AQFT program.)
\item We might bite the bullet and accept that covariance is only an
approximation.  This would follow, for instance, if X were itself a
non-covariant theory (such as Barbour (\citeyearNP{barbour1,barbour2,barbour99}) has advocated in quantum
gravity).
\item We might find that X is itself a totally covariant theory.  There
seems no reason why such a theory should not have a family of useful non-
covariant approximations.
\item Perhaps most interestingly, X may be a theory which does not
involve spacetime at all at a fundamental level (as is conjectured would be the
case in loop-space quantum gravity, or non-perturbative string theory).
In this case, relativistically covariant spacetimes would have to emerge
as effective, approximate concepts from such a theory, and it would seem
to be a curiosity rather than a pathology if this emergence proceded
via non-covariant intermediate theories.  After all, when we say that we
expect spacetime to be quantized, we generally do not mean that it is
literally to be divided into grid squares, but rather that it is to be
replaced in toto with some fundamentally discrete entity which in some
circumstances is approximately isomorphic to spacetime --- yet in
recovering that spacetime it might be necessary to use an intermediate
step which does involve treating spacetime as a grid.  (To take an
analogy, we think that the phase space of classical mechanics breaks
down at scales of $\sim \hbar$, but that doesn't mean that we think that
phase space really is a grid of $\hbar$-sized squares.)
\end{enumerate}

\section{Inequivalent representations}\label{represent}

In section \ref{observables}, we defined algebraic QFTs in terms of a
net of \emph{operator} algebras.  The usual definition, however, is of a
net of \emph{abstract} algebras, which can be represented as Hilbert
space operators in a number of inequivalent ways.  At first sight these
inequivalent representations seem to cause problems for QFT since there
seems no principled way to select the `right' one; it is also not
immediately obvious what has happened to the inequivalent
representations in our previous quantization (in section
\ref{wavefunctional}) of QFT in terms of wave-functionals on
configuration space.  In this section we shall address these issues, and
will show that inequivalent representations --- although important and
interesting in QFT --- pose no problems for its foundations.

\subsection{The algebraic approach}\label{repint}

Consider the quantum mechanics of some finite number of nonrelativistic, scalar particles.  
We could specify this theory by starting with the Hilbert space, and then describing the 
actions of the various observables (position, momentum and functions thereof) on the space.  
We would need to specify a specific function of the observables as the Hamiltonian, 
so as to tell us how states evolve in time.

Once we have these observables, we could shift to the Heisenberg representation by 
using the Hamiltonian to time-evolve the observables:
\[\op{X}(t)= \exp(-i t \op{H}) \op{X} \exp(+i t \op{H}).\]
(The Hamiltonian would then be unnecessary, since all the information about the system's 
time evolution would be contained within the time-evolved observables.)
Let \mc{A} be the algebra of these observables.\footnote{If we were being
mathematically rigorous, we would instead work with the algebra of
\emph{bounded functions} of the observables, since \op{X} and \op{P} are
unbounded and thus very awkward to work with.}


Now suppose we take $\mc{A}$ considered as an abstract algebra (with the norm topology, 
\ie a $C^*$-algebra) and discard the Hilbert space.  We would still have a great deal of 
dynamical information encoded in the algebraic and topological structure of the algebra, 
but would have lost the action of the algebra on physical states.

Suppose we try to recover the Hilbert space action.  That is, we wish to find an isomorphism 
between the abstract algebra $\mc{A}$ and some subalgebra of the bounded operators on a 
Hilbert space \hilbert{H}, \ie a \emph{representation}\footnote{Technically, a 
representation need not be an isomorphism.  However, here we are working with 
simple $C^*$-algebras, for which all irreducible representations are faithful 
(\ie isomorphic to the original algebra).} of $\mc{A}$.  If this 
representation turns out to be unique up to unitary equivalence (an assumption 
which was often made uncritically in the early days of quantum theory) then we 
must have recovered the original Hilbert space, \ie the representation must be isomorphic
to the original choice of Hilbert space and of observables on it.

In fact, for a system of finitely many particles the representation is indeed unique --- provided 
we require it to be \emph{irreducible}.  This last is the requirement that the 
Hilbert space does not split into orthogonal subspaces preserved by the action of 
$\mc{A}$.  Equivalently, there must exist in the Hilbert space a \emph{cyclic vector} 
whose image under the action of $\mc{A}$ is dense in the Hilbert space.  Physically 
this requirement rules out the existence of superselection rules: superselection 
sectors can be identified with the irreducible sectors of a reducible representation.

By an \emph{algebraic} quantum theory, we shall mean one which is specified by giving 
the algebraic structure of the observables but not the Hilbert space on which they act.  
A theorem of Stone and Von Neumann tells us that any such quantum theory with a 
\emph{finite} number of degrees of freedom --- \ie finitely many `position' and 
`momentum' observables, corresponding to a classical theory with a finite-dimensional 
phase space ---  has a unique irreducible representation (up to unitary equivalence).  
Hence for such systems the algebraic and Hilbert-space ways of specification are equivalent.  

\subsection{Infinite-dimensional systems}

When the number of degrees of freedom of a theory becomes infinite, the Stone-von Neumann theorem 
fails.  Such theories generally have a large number of inequivalent
representations, and so giving the algebraic structure does not
completely specify the theory.  

Nonetheless it still seems natural to specify a theory algebraically, because
\begin{enumerate}
\item The way in which we usually describe a system is to describe its
observables.  Furthermore, a theorem of Fell (see below) tells us that
no amount of empirical data can tell us what the `real' representation is.
\item As explained in section \ref{observables}, the `quantization'
process begins with the classical observables and tries to map them onto 
quantum-mechanical operators.  Since in this process we are trying to
impose purely algebraic restrictions on this map (specifically, the restriction
that Poisson brackets go over to commutators) it will generally specify only the
algebraic structure of a quantum theory, with an ambiguity left as to
the Hilbert space action.  In other words, two quantum theories which
are algebraically identical could be regarded as equally valid
quantizations of the classical system.\footnote{Separately from this, of 
course, we might not in general expect quantization even to produce a unique
\emph{algebraic} structure for the quantum observables, because of ambiguities
about operator ordering.}
\end{enumerate}

If we were to take a robustly instrumentalist viewpoint there would be no
particular problem: Fell's theorem says that a finite number of
measurements, each conducted with finite accuracy, cannot distinguish
between representations.  Hence we could reproduce any experimental
results using whichever representation was most convenient.\footnote{It
is important in this approach that we have all the observables in the
algebra.  If, for instance, the (renormalized) stress-energy tensor is
not in the algebra, then its values may distinguish
representations \cite{waldqft}.}

From a realist standpoint, however, we appear to have a dilemma.  Even
if no empirical data lets us distinguish between representations,
nonetheless there should be a fact of the matter as to which is the
`correct' representation; yet it is provably impossible for us ever to
discover this fact.  To see this, suppose we have two
representations on Hilbert spaces $\hilbert{H}_1$\ and $\hilbert{H}_2$,
and suppose the `real' representation is the first.  Then if the system
is in some state \ket{A}, Fell's theorem tells us that 
\begin{quote}
For any operation $\mc{O}$ carried out with finite accuracy on \ket{A},
there is a state \ket{\mc{O};\mc{A}} in $\hilbert{H}_2$ such that all data
resulting from the operation are consistent with the real representation
being on $\hilbert{H}_2$ and the real state being \ket{\mc{O};\mc{A}}.
\end{quote}

In fact, this dilemma will prove only apparent, as we will find in the
next three sections when we study representation ambiguities in more
detail.  To do so, we distinguish two ways in which inequivalent
representations can occur: one associated with the short-distance and
high-energy
(`ultra-violet') degrees of freedom, one with the long-distance 
(`infra-red') ones.  Given our tolerant attitude to short-distance cutoffs
(c.\,f.\, section \ref{renorm}), it should come as no surprise that all the
important sources of inequivalent representations of a given QFT fall into the second category: 
for the Stone-von Neumann theorem guarantees uniqueness
of representation for any theory with finitely many degrees of freedom,
and a field theory in any finite region has only finitely many degrees
of freedom below a finite energy bound.  We will argue that the
ultra-violet representation ambiguity is a pure mathematical artefact,
whilst the infra-red ambiguity is physically real but non-pathological.

\subsection{Ultra-violet degrees of freedom}

It is probably already clear to the reader why we can ignore the 
ultra-violet representation ambiguity: UV-inequivalent 
representations occur because of the existence of degrees of freedom at
arbitrarily short lengthscales, and in section \ref{renorm} it was
argued that real QFTs are cut off at short, though finite lengthscales.
In particular, the discretised QFTs of section \ref{cutoffconcept} have
only finitely many degrees of freedom per space-time point, and hence no
UV-inequivalent representations.  Since renormalisation theory
tells us that any QFT is scale-$l$ equivalent to some such discretised QFT
(on lengthscales long compared to the grid size for the latter theory),
it follows that any occurence of UV-inequivalent representations is
purely a mathematical artefact and has no physical significance.  

\subsection{Inequivalent representations in non-relativistic physics}

The physical significance of IR-inequivalent representations has long
been appreciated in AQFT; it will be reviewed here from the perspective
which this paper adopts towards QFT, but with no claim to originality.
We begin by returning to non-relativistic QM: imagine a line of $n$ two-state systems --- 
spin-half particles fixed in place, say.  The Hilbert space of each system is two-dimensional, 
and we can construct the overall Hilbert space by taking the $n$-fold tensor product.  
The resulting space will have dimension $2^n$, which is as we expect since we must make 
$n$ yes-no choices to choose a state with (say) given spin in the z-direction for each component
system.

Now suppose our system becomes infinitely large.  The dimension of the system will be infinite, 
of course, but it will be a larger infinity than those to which we are used --- specifically 
it will be $2^{\aleph_0}$, the cardinality of the continuum, which is strictly larger than 
the cardinality $\aleph_0$ of the integers.\footnote{See, \egc, \citeN{enderton}.}

It follows that systems with infinitely many components have a Hilbert space which is 
non-separable (\iec, has uncountable dimension).  To see the consequences of this, consider 
the operator algebra of our set of two-state systems.  It consists of the set of 
linear combinations of spin operators, and hence has countably many linearly independent elements. 
The action of this algebra on any given state will generate only countably many linearly independent 
states, hence the action of the operator algebra on the total, non-separable space is highly 
reducible.

In other words, the non-separable space has decomposed into various superselection sectors 
(uncountably many, in fact), each giving an irreducible representation of the operator algebra.  

To see what these sectors are, suppose we start with all components having spin up.  Then 
the action of any element of the algebra can, at most, cause finitely many components to have spin down.  
So no amount of algebraic action can transform such a state into one in which, say, every 
second component has spin up.  This state, in turn, can be transformed into other states 
differing from it in finitely many places, but not into a state in which all components are 
spin down \ldots or every third component is spin down \ldots or where half the states are 
spin up but the spin-up states are grouped in pairs \ldots

In other words, the different representations describe states which are  `infinitely' 
different from one another (see \citeN{morrison} for a full discussion).  In statistical 
mechanics, different representations describe different phases, since at arbitrarily large 
length-scales --- tending to the thermodynamic limit --- only the infinite differences between 
representations remain visible.

For such systems, of course, the use of different representations is an idealization, since 
the actual systems are finite.

\subsection{IR-inequivalence in quantum field theory}

Guided by the nonrelativistic example, we look for inequivalent representations in 
field theory by looking for states differing on asymptotically large scales.  The 
difference from the previous section is that here infinite systems are
perfectly respectable: in fact, our discretised QFTs are effectively
infinite crystals.

This resolves the problem mentioned at the start of section
\ref{represent}: what is the connection between inequivalent
representations and the wave-functional approach of section 
\ref{wavefunctional}?  The answer is that the space over which the 
wave-functionals are defined --- the field-configuration-space \mc{S} 
--- is infinite-dimensional both because functions may vary on
arbitrarily short length-scales, and because they may have arbitrary
large-distance boundary conditions.  The former reason is nullified when
a short-distance cutoff is imposed (as in section \ref{renorm}) but the
latter one also interferes with the definition of the functional
integral, and has to be dealt with by imposing boundary conditions at
infinity (such as the requirement that functions in \mc{S} be square-
integrable).    Each choice of boundary condition generates a different,
and inequivalent, representation.

Examples of inequivalent representations are then:

\subsubsection*{Differing mass-densities at infinity}

The closest analogue to the previous section is the case of states with asymptotically 
nonvanishing mass density: average density is superselected \cite{haag} and different 
densities correspond to different representations.  It is worth noting that this rules 
out the Fock representation for an open universe, even a flat one, since any realistic state 
of such a universe will have nonzero mass density; the Fock representation, which 
has finite-energy states, stands in the same relation to a realistic open-universe 
representation as an asymptotically empty classical spacetime does to a realistic open 
cosmology \cite{wald}.

The overall velocity of the universe is also superselected, of course, except in the 
special case where the (expected value of the) stress-energy tensor is asymptotically diagonal.

\subsubsection*{Differing total charge}

Even when the spacetime is asymptotically empty, inequivalent representations exist.  
The total charge of a system leaves its imprint on the spacetime at arbitrarily long 
distances in the form of the asymptotic Coulomb field, whose flux through a sphere is 
independent of the size of that sphere.  As such, total charge is superselected \cite{strocchi}.  
(Another way to see this is to remember that field lines extend to infinity, so no amount of 
local operations will remove them.)

In fact, the centre-of-mass velocity of a charged system is also superselected \cite{haag}, 
since the asymptotic field will be a Lorentz-boosted Coulomb field and the associated magnetic 
field lines also extend to infinity.

\subsubsection*{Inequivalent vacua}

A major component of the Standard Model of particle physics \cite{peskinschroeder} is the presence of 
field theories with degenerate vacua.  Classically this corresponds to the existence of nonzero 
field configurations which are global minima of the energy.  If the field theory has a continuous 
global symmetry then there will exist a continuum of vacuum states, wave-packets around which will 
give us the quantum ground states.  Since these states differ from one another everywhere in space, 
they too belong to inequivalent representations.

\subsubsection*{Changes in the interaction parameters}

A famous theorem (Haag's theorem; see \citeNP{haag}) tells us, in effect, that interacting fields 
cannot be represented on the same Hilbert space as the corresponding free fields, even when we 
consider asymptotically early or late field operators.  In the current framework this may be 
understood as follows: introducing the interaction will change the ground state everywhere in 
space, causing it to be infinitely different from the free-field ground state and so in a different 
superselection sector.

\vspace{0.5cm}

In fact, and as these examples show, there is not really anything
`quantum' about IR-inequivalent representations: in many cases, picking
a representation is rather like picking boundary conditions at infinity in a
classical problem.  The following classical --- and utterly banal ---
version of Fell's theorem shows why the `dilemma for realism' mentioned
in section \ref{repint} is nothing of the kind as far as IR-inequivalent
representations are concerned:
\begin{quote}
\textbf{Classical Fell `theorem':} the spatial average of a classical quantity
over
an infinite universe cannot be empirically determined by measurements
confined to a finite spatial region.
\end{quote}

\subsection{Retrospective}

From a realist perspective the sting of the representation ambiguity has largely been drawn.  
Locally, any representation ambiguity is artificial, caused by the presence of unphysical 
degrees of freedom beyond the high-energy limit of the theory's validity.  Globally, there 
may indeed be representation ambiguities --- depending on cosmology, and the topology of 
the universe --- but the inaccessible information which they encode is `respectable', 
analogous to the classical inaccessibility of the long-distance structure of the universe.

Nonetheless it would be more aesthetic to lose the long-distance ambiguity as well.  
This can only be done if the universe is finite.  There has long been a divide between 
observational cosmology (favouring an open universe) and theory (preferring closed universes); 
this analysis of representations makes a modest contribution to keeping the divide open.

For the purposes of this paper, representational ambiguities need no longer concern us.  
In practice we are always analysing a theory in a finite region --- and idealizing the system 
beyond that region in whatever manner is convenient --- so different choices of representation 
should not affect our conclusions.  

\section{Localisation in quantum field theory}\label{local}

The phenomenology of particle physics --- and indeed, of virtually all
of science --- makes extensive use of the concept of localisation: that
is, of the concept that physical systems have at least some states which localised in finite
spatial regions.  There are a number of results in AQFT which apparently
rule out the possibility of such states, and this is sometimes described
as a paradox in QFT.  In this section the problem of localisation will
be analysed from the viewpoint of the previous sections, and it will be
argued that nothing paradoxical is going on.

In a relativistic theory, there are two natural ways of thinking about
localisation: \emph{spatial} localisation, where something is
localised in some subregion $\Sigma_i$ of a spacelike slice $\Sigma$
(and therefore \emph{not} localised anywhere else in $\Sigma$), and
\emph{spacetime} localisation, where something is localised in some
subregion \mc{O} of spacetime (and therefore \emph{not} localised in any
region of spacetime spacelike separated from \mc{O}).  The latter
concept makes sense in AQFT (including the scale-$l$ approximate AQFTs
of section \ref{cutoffconcept}); the former needs some concrete QFT to be
interpreted.  In this section we will move freely between the two
notions, using a discretised QFT where necessary to make sense of
spatial localisation, and relying on renormalisation theory to ensure
that we are not making cutoff-dependent statements.

\subsection{Localisation and the Reeh-Schlieder theorem}\label{localreeh}

We have already established (in section \ref{observables}) that in QFT
the idea of localisation arises through the spatial localisation of the
field observables $\op{\phi}(\vctr{x})$ and $\op{\pi}(\vctr{x})$ (here,
for convenience, we work with a concrete element of the scale-$l$
equivalence class of the QFT in question).  But how are we to go from
localised observables to localised states?  We might begin by trying the following
(phrased in terms of spacetime localisation):

\begin{quote}\textbf{Naive localisation}: A state \ket{\psi} is localised in a
region \mc{O} iff $\matel{\psi}{\op{A}}{\psi}=0$ for any observable
$\op{A}$ localised in a region spacelike separated from \mc{O}.
\end{quote}

This seems plausible when we compare to the classical case: there a
state is localised in \mc{O} if $\pi(x)=\phi(x)=0$ for any
$x$ spacelike separated from \mc{O}.  But it is mathematically impossible for any states to satisfy 
it, for it implies that for any such $x=(\vctr{x},t)$, and for any $n$, 
\be
\matel{\psi}{\op{\phi}^n(\vctr{x},t)}{\psi}=\matel{\psi}{\op{\pi}^n(\vctr{x},t)}{\psi}=0.\ee
But this would imply that \ket{\phi} was a simultaneous eigenstate of
$\op{\pi}(\vctr{x})$ and $\op{\phi}(\vctr{x})$, and these operators have
no eigenstates in common (the mathematics, once we have introduced a
cut-off
to deal with operators defined at a point, is the same as for the
nonrelativistic operators $\op{X}, \op{P}$, which are well-known to have
no eigenstates in common).

Physically it is easy to see what is happening here.  The vacuum state of a
field theory (which we will denote by \ket{\Omega}) is not `nothingness', or `empty space'; it is simply a
slightly colourful way of describing the ground state of the field's
Hamiltonian.  In solid-state systems
this state is just the zero-temperature state of the solid, in which the
atoms will not be at rest but will have zero-temperature fluctuations;
the same will be true for the field excitations of a relativistic field
theory.  

This suggests, however, an alternative: rather than consider states
which are local \textit{simpliciter}, we could consider states which
\emph{differ} only locally.  The vacuum could be used as a reference state, and
states could be declared `local' if they differed only locally from the
vacuum.  

To make this quantitative, we need to understand what it could mean for
two quantum states to `differ locally'.  Recall that in section
\ref{wavefunctional} we defined spatially local subsystems of the QFT,
with Hilbert spaces $\mc{H}_{\Sigma_j}$, which described the degrees of
freedom of the QFT localised in $\Sigma_j$; this formally defined
concept can be exactly defined in discretised QFT, provided that the
grid spacing is small compared to each $\Sigma_j$.  If the vacuum were a
product state with respect to any decomposition of the form
$\mc{H}_{\Sigma}= \mc{H}_{\Sigma_1}\otimes \mc{H}_{\Sigma_2}$ (where $\Sigma = \Sigma_1 \cup 
\Sigma_2$ and $\Sigma_1\cap\Sigma_2=\emptyset$)\footnote{Since we are working with discretised QFT 
here, a more accurate statement would be `all grid points in $\Sigma$ are
contained either in $\Sigma_1$ or $\Sigma_2$, and $\Sigma_1$ and
$\Sigma_2$ do not overlap.'} then defining `differing
locally' would be straightforward: the vacuum would have form
$\ket{\Omega}=\ket{\Omega_1}\otimes\ket{\Omega_2}$, and any state
\ket{\psi} could be said to differ from the vacuum only in $\Sigma_1$ if
(a) \ket{\psi} was also a product state, $\ket{\psi}=\ket{\psi_1}\otimes
\ket{\psi_2}$, and (b) $\ket{\psi_2}=\ket{\Omega_2}$.

However, the vacuum is not a product state, but a highly entangled one (this can be readily seen by
calculating expectation values such as 
\be\matel{\Omega}{\op{\phi}(x) \op{\phi}(y)}{\Omega} - \matel{\Omega}{\op{\phi}(x)}{\Omega} 
\matel{\Omega}{\op{\phi}(y)}{\Omega}\ee
and showing that they are non-zero for all spacelike separated $x,y$,\footnote{For a free scalar 
field, with $x=(\vctr{x},t)$ 
and $y=(\vctr{y},t)$ we get \[\matel{\Omega}{\op{\phi}(x) \op{\phi}(y)}{\Omega} - \matel{\Omega}{\op{\phi}(x)}{\Omega} 
\matel{\Omega}{\op{\phi}(y)}{\Omega}=\frac{1}{2}(m^2-\nabla^2)^{-\frac{1}{2}}\delta(\vctr{x}-\vctr{y});\] smearing $\op{\phi}(\vctr{x},t)$ and $\op{\phi}(\vctr{x},t)$
out with non-overlapping spatial test-functions $f,g$, this becomes
$\frac{1}{2}\int_\Sigma f(\vctr{x}) [(m^2-\nabla^2)^{-\frac{1}{2}}
g](\vctr{x})$.  That this is nonzero for generic $f,g$, irrespective of
the spatial separation of their supports, follows from the antilocality
of the operator $(m^2-\nabla^2)^{-\frac{1}{2}}$ \cite{segal}.  Adding interactions to the free field
will lead to perturbative modifications to these results but should not change their 
qualitative nature, provided 
the interactions are weak enough at large distance-scales to treat perturbatively --- hence 
adding a 
$\lambda \phi^4$ interaction will have only quantitative effect, but adding a coupling 
to a non-Abelian gauge
field may have more drastic consequences.} and is discussed by
\citeN{clifton}.)  As such, defining `differing locally' is somewhat more
subtle, since it is difficult and even controversial to say what the
state of a subsystem of a quantum system is, when the total state of the
system is entangled.  Some possible definitions might be:
\begin{enumerate}
\item Two states might be said to differ only within $\Sigma_1$ if their
expectation values coincide with respect to all operators localised
outside $\Sigma_1$.  
\item We might take the `state of a subsystem' to be the density operator obtained
when all degrees of freedom of the other subsystems are traced over.  In this case, two states \ket{\psi_a} and \ket{\psi_b}
could be said to differ only within $\Sigma_1$ if (with $\tr_{\mc{H}_{\Sigma_1}}$ being 
tracing over degrees of freedom inside $\Sigma_1$)
\be
\tr_{\mc{H}_{\Sigma_1}}\proj{\psi_a}{\psi_a}=\tr_{\mc{H}_{\Sigma_1}}\proj{\psi_b}{\psi_b}.\ee
\item The problem with the above definition of the state of a subsystem
is that we cannot then recover the state of the whole system from the
states of its components.  This could be taken to indicate the intrinsic
nonlocality of quantum states; this view was recently rejected by
\citeN{deutschhayden}, who propose that locality in quantum mechanics is
best understood in the Heisenberg picture.  They pick a specific
Hilbert-space 
vector as a reference state, to be used to calculate all expectation values
irrespective of the particular initial conditions and then treat the quantum state as being
specified at time $t$ by the set of all  Heisenberg operators at time
$t$.  The obvious definition of `differing locally' within this
framework is that two states differ only within $\Sigma_1$ if they have
the same Heisenberg operators outside $\Sigma_1$.
\item Leaving aside the question of how to specify the state of a
subsystem, we might define two states as differing only within $\Sigma_1$ if they are
connected by a unitary operator localised within $\Sigma_1$ (such an operator
will have form $\op{U}_{\Sigma_1}\otimes\id_{\Sigma_2}$ acting on the tensor-product
space $\mc{H}_\Sigma=\mc{H}_{\Sigma_1}\otimes\mc{H}_{\Sigma_2}$).
\end{enumerate}

Fortunately, all of these definitions coincide\footnote{The equivalence of (1), (2) and
(4) is trivial; the equivalence of (3) and (4) is given in \citeNP{deutschhayden} (and is, again,
trivial once Deutsch and Hayden's formalism is understood).} and this in turn
supports the naturalness of each of them as a definition of `differing
locally'.  Using the vacuum as our reference state, and using for convenience the first
definition, we obtain our definition of localised states, which we refer
to as Knight localisation after \citeN{knight} who first proposed such a
definition.  We state it in the more general context of AQFT (for which reason we
return to spacetime, as opposed to spatial, localisation):

\begin{quote}\textbf{Knight localisation:} a state \ket{\psi} is localised in
a region \mc{O} iff 
$\matel{\psi}{\op{A}}{\psi}- \matel{\Omega}{\op{A}}{\Omega}=0$ for any observable
$\op{A}$ localised in a region spacelike separated from \mc{O}.
\end{quote}

However, Knight localisation differs in one important respect from the
sort of localisation which we encounter in NRQM.  In the latter,
properties like `is localised in \mc{O}' are treatable in the same way as properties
like `has energy $E$' or `has momentum less than $p$': that is, we can define a 
projection operator whose intended interpretation is `localised in
\mc{O}', whose range is the space of all such states.  This would be possible for 
Knight-localised states iff
they form a subspace: that is, iff any superposition of two states
Knight-localised in \mc{O} is also Knight-localised in \mc{O}.  

The fact that Knight-localised states do not have this property is a
consequence of the Reeh-Schlieder
theorem \cite{reehschlieder}.
\begin{quote}\textbf{Reeh-Schlieder theorem}: for any region \mc{O},
the set of vectors $\mc{A}(\mc{O})\ket{\Omega}$ generated by the action
of operators localised within \mc{O} upon the vacuum, spans the Hilbert
space of the QFT.
\end{quote}
(For a proof, and further discussion, see \citeNP{haag}.)
It follows\footnote{To see that it follows, we need only note that the unitary 
elements of a (bounded) operator algebra $\mc{A}(\mc{O})$ span $\mc{A}(\mc{O})$.  This can be proved as 
follows: for any bounded Hermitian element $\op{H}$ of $\mc{A}(\mc{O})$, and any $t\neq 0$, $(it)^{-1}(\exp(i
t \op{H}) - \id)$ is a linear combination of unitary elements of $\mc{A}(\mc{O})$.  As
$t\rightarrow 0$, this sequence tends to \op{H}, hence \op{H} is in the
span of the unitary operators.  To complete the proof, simply recall
that any linear operator can be written as $\op{A}+ i \op{B}$, where
\op{A} and \op{B} are Hermitian.}
from the Reeh-Schlieder theorem that states
Knight-localised at \mc{O} span the entire state space, which rules out
any possibility of a projector meaning `localised with certainty in \mc{O}'.

It has long been understood that the Reeh-Schlieder theorem is a
consequence of the entanglement of the QFT vacuum.  Non-relativistic
quantum mechanics furnishes us with many examples where, given an entangled state 
of the Hilbert space $\mc{H}_A \otimes \mc{H}_B$, unitary operations on
$\mc{H}_A$ alone suffice to produce a basis for $\mc{H}_A \otimes
\mc{H}_B$.  One example, which plays an important role in quantum teleportation \cite{bennett},
is the \emph{Bell basis}: we take a system of two qubits (\iec, two-state systems) 
and prepare them in one of these four states 
\be
\begin{array}{rcl}
\ket{B1} & = & \nrm \left( \tpk{0}{0} + \tpk{1}{1} \right) \\
\ket{B2} & = & \nrm \left( \tpk{0}{0} - \tpk{1}{1} \right) \\
\ket{B3} & = & \nrm \left( \tpk{1}{0} + \tpk{0}{1} \right) \\
\ket{B4} & = & \nrm \left( \tpk{1}{0} - \tpk{0}{1} \right) \end{array}.
\ee
Obviously this is a basis for the combined two-qubit system, but also any element of it is cyclic under unitary operations carried out on the first qubit alone, as is easily shown:
\be
\begin{array}{rcl}
(M_A \otimes \id) \ket{B1} & = & \ket{B2} \\
(M_B \otimes \id) \ket{B1} & = & \ket{B3} \\
(M_C \otimes \id) \ket{B1} & = & \ket{B4} \end{array}
\ee
where
\[
M_A = \mtr{1}{0}{0}{-1}; M_B = \mtr{0}{1}{1}{0}; M_C = \mtr{0}{1}{-1}{0}\]
in the $(\ket{0},\ket{1})$ basis.  In other words, even if the qubits are macroscopically separated, 
if they begin in an appropriately entangled state then it is possible by operating on one qubit 
to generate a set of states which span the Hilbert space.  (Examples of
this kind are analysed in rather more detail by \citeN{redhead} and \citeN{clifton}.)

Since (as discussed above) the QFT vacuum is entangled, the 
Reeh-Schlieder theorem should come as no surprise to us.  Nor does it
cause any \emph{logical} problems for Knight's definition of
localisation: there are many perfectly respectable properties of quantum
states which are not preserved under linear superposition, such as being
an eigenstate of energy, or being an entangled state.

It does, however, cause us \emph{practical} problems.  The sort of
localisation which we use in NRQM \emph{is} preserved under linear
superpositions, and this fact is essential to the analysis of NRQM
problems --- so, at least on the face of it, NRQM localisation and
Knight localisation must be different concepts.  Nor is the problem
confined to NRQM: in  scattering theory, for instance, it is crucial to
be able to discuss the amplitude for a particle to be scattered to a
given area, and such discussions presuppose that spatial localisation is
preserved under superpositions.  

So what is going on?  In the face of the Reeh-Schlieder theorem, there
is little prospect of finding an alternative to Knight localisation
which does have the required properties (see \citeN{halvorsonclifton} for
a wide variety of no-go theorems); so, how are NRQM, and scattering
phenomenology, compatible with QFT?

The key to this question is the fact that NRQM is not supposed to be
\emph{perfectly} compatible with QFT: rather, it is supposed to be an
approximation to QFT valid only in certain regimes of QFT.  (And the
same is true for scattering theory, which is a marvellous tool to
analyse high-energy collisions, but of limited use in understanding,
\egc, quark confinement.)  With this in mind, we will lower our sights and
seek an approximate form of locality which, in certain domains of QFT,
will be approximately preserved under linear superposition.  Section
\ref{effloc}
will construct such an approximation, and section \ref{loccon} will address the
question of whether it is after all adequate for our needs.

\subsection{Effective localisation}\label{effloc}

Our approximate concept of localisation will be defined as follows (it is 
again defined in terms of spatial localisation, partly for ease of comparison
with nonrelativistic localisation):

\begin{quote}\textbf{1. Effective localisation (qualitative form):} A state \ket{\psi} is 
\emph{effectively localised} 
in a spatial region $\Sigma_i$ iff for any function $\op{f}$ of field operators
$\op{\phi}, \op{\pi}$,
$\matel{\psi}{\op{f}}{\psi}- \matel{\Omega}{\op{f}}{\Omega}$
is negligibly small when $\op{f}$ is evaluated for field operators outside $\Sigma_i$, 
compared to its values when evaluated for field operators within $\Sigma_i$.
\end{quote}

\begin{quote}\textbf{2. The effective localisation principle (ELP) (qualitative form}: A
subspace \mc{H} of the QFT Hilbert space $\mc{H}_\Sigma$ obeys the ELP
on scale $L$ iff for any spatial region \mc{S} large compared with $L$, a superposition
of states effectively localised in \mc{S} is effectively localised in effectively the same 
region.
\end{quote}

These qualitative notions can be made precise in a number of ways, such as:
\begin{quote}\textbf{1. Effective Localisation (quantitative form):} A state is 
\emph{L-localised} in a region $\Sigma_i$, iff for any function $\op{f}$ of field operators
$\op{\phi}, \op{\pi}$,
$\matel{\psi}{\op{f}}{\psi}- \matel{\Omega}{\op{f}}{\Omega}$
falls off for large $d$ like (or faster than) $\exp(-d/L)$, where $d$ is the distance 
from $\Sigma_i$
at which the function $\op(f)$ is evaluated.  (Note that there is no difference, 
according to this definition, between a state $L$-localised at some spatial point 
\vctr{x} and a state $L$-localised in a region of size $\sim L$ around \vctr{x}.)
\end{quote}

\begin{quote}\textbf{2.  ELP (quantitative form):} A state obeys the ELP on scale 
$L$ iff, for any 3-sphere \mc{S} of radius $>L$, a superposition of states $L$-localised 
in \mc{S} is $L$-localised in \mc{S}.
\end{quote}

Obviously ELP cannot hold on any scale for the Hilbert space $\mc{H}_\Sigma$ of the
full QFT, on pain of violating the Reeh-Schlieder theorem (the space of states effectively
localised in a region obviously includes all those Knight-localised in a region, and as mentioned
in section \ref{localreeh}, the span of all such states is the whole of $\mc{H}_\Sigma$.)
However, if some subspace \mc{H} of $\mc{H}_\Sigma$ is such that:
\begin{enumerate}
\item ELP holds, for all regions large compared with some lengthscale $L$;
\item the current state of the QFT lies within \mc{H}; and
\item \mc{H} is approximately conserved, on timescales of interest
to us, by the QFT dynamics
\end{enumerate}
then we will, approximately, be able to define projection operators in
\mc{H} which project onto states localised in a given region, provided
that that region is large compared to $L$.
(Note that $L$ is nothing to do with the cutoff lengthscale $l$, except
that $L$ must be large relative to $l$ for our results to be 
cutoff-independent.)

Why should we expect ELP to hold for any subspaces of $\mc{H}_\Sigma$?
To answer this, we need to analyse the entanglement of the vacuum in somewhat more detail.
For if it were completely non-entangled, ELP would indeed hold for all
of $\mc{H}_\Sigma$, as argued for above.  We might expect, then, that if
the vacuum entanglement drops off rapidly above certain distance-scales,
then above those scales localisation would be `almost' preserved under
superposition, and only extremely careful superposition of states would
spoil this; hence, we would expect it to be fairly typical for a given subspace 
of $\mc{H}_\Sigma$ to satisfy ELP on these lengthscales.

Does vacuum entanglement in fact decrease with distance; equivalently, do the 
correlations in the vacuum  decrease with distance?  For a massive QFT it can be
shown that they drop off exponentially, with the lengthscale given by
the `Compton wavelength' $1/m$ (where $m$ is the renormalised mass term in the QFT Hamiltonian):
this is provable rigorously for any QFT satisfying the AQFT axioms
\cite{fredenhagen}, and can be calculated perturbatively in the case of
QFTs with weak large-scale interaction terms.  (QCD, and other theories
with strong long-distance interactions, may not be covered by either
case --- rather little is known about the QCD vacuum, in fact.)

To see intuitively the significance of the $1/m$ lengthscale, recall
that the Hamiltonian for a free scalar QFT has the form
\be\op{H} \Psi[\chi]= \frac{1}{2}\int_\Sigma \dr{^3\vctr{x}} 
\left( \frac{\delta^2}{\delta \chi(\vctr{x})^2} + (\nabla \chi)(\vctr{x})^2 + m^2 \chi(\vctr{x})^2 
\right) \Psi[\chi],\ee 
and that in a discretised QFT we replace derivative terms like $\nabla
\op{\phi}(\vctr{x})$ with terms like $(\phi(\vctr{x})-
\phi(\vctr{x}+\vctr{l}))/l$, 
where $|\vctr{l}|=l$ and $l$ is the cutoff length.  These discretised
derivative terms are the only ones in the Hamiltonian which lead to
entanglement, as they are the only ones which couple observables at different lattice points.

Now there is no
requirement that $l \ll 1/m$ --- all we require of $l$ is that it is
small compared with the lengthscales at which we want to study the QFT's
structure.  Admittedly, if $l$ is smaller than, or comparable to, $1/m$, then the
coupling terms $(\phi(\vctr{x})-
\phi(\vctr{x}+\vctr{l}))/l$, which are responsible for entanglement,
will be significant relative to the other terms in the Hamiltonian.
However, as $l$ becomes large compared with $1/m$, the coupling terms
become only a small perturbation on the Hamiltonian
\be\op{H} \Psi[\chi]= \frac{1}{2}\int_\Sigma \dr{^3\vctr{x}} 
\left( \frac{\delta^2}{\delta \chi(\vctr{x})^2} +  m^2 \chi(\vctr{x})^2 
\right) \Psi[\chi],\ee 
which is the Hamiltonian of a set of uncoupled oscillators, one at each
lattice point.  Thus, on such scales we should expect to find the ground
state only slightly entangled.

So, we expect that for a massive QFT with mass $m$, we should find it to be common for the ELP 
to hold for subspaces on lengthscales large relative to $1/m$.  Obviously, this heuristic
argument has to be tested for any specific subspace in which we are interested.  In
particular, we find it to  hold for the one-particle sector of any free (or asymptotically free)
massive bosonic QFT, suggesting that such theories have a quite satisfactory notion of particle
localisation (this result is proved, and analysed \textit{in extenso}, in 
\citeN{wallaceparticle}).

For a QFT which describes some massless fields, the vacuum correlation functions
still drop off with distance, but follow a power-law dropoff rather than an
exponential one; hence there is no characteristic lengthscale on which
correlations exist.  This does not rule out finding subspaces of
the Hilbert space of the QFT for which ELP applies on some useful scale;
however, it suggests that the scale will vary according to the specific
problem being examined.

Furthermore, we expect there to be physically relevant sectors of at
least some massless QFTs in which states are localised only on extremely
large lengthscales (if at all).  For instance, there should be a 
low-energy sector of QED in which electron number is approximately conserved, in which
the electromagnetic field has effectively no independent degrees of
freedom, and in which each electron has an associated Coulomb field.  In
such sectors, the state of the electromagnetic field is not really
localised at all, or at least not on any characteristic lengthscale;
analysis of such sectors would be necessary to establish the validity of
ELP for nonrelativistic, interacting particles. (See chapter VI of
\citeN{haag} and references therein for further discussion of the state
space of QED, from the perspective of AQFT.)

\subsection{The conceptual justification of effective
localisation}\label{loccon}

We have now constructed a definition of localisation which approximately
has the properties we wish: that is, it designates a set of states which
are approximately localised, and that set is closed under linear
superpositions.  But is this approximate localisation good enough?
After all, in general `effectively localised' states are not exactly
localised --- particle states, in particular, invariably differ from the
vacuum everywhere in space.  It follows that particle creation and
annihilation operators cannot be exactly localised, which seems to
preclude localised particle-detecting devices.

This problems has recently been addressed from an AQFT viewpoint by
\citeN{halvorsonclifton} (see also \citeN{haag} for a more technical
discussion along similar lines).  Their argument is as follows:
\begin{enumerate}
\item What we actually measure are (exactly) local operators.
\item Although particles cannot be detected by measurements of any local
operator, they can be detected by measurements of operators which are
very close (in operator norm) to local operators.
\item It follows that to a high degree of accuracy, we \emph{can}
detect particles; however, the detection will not be 100\% reliable.
\end{enumerate}
In their approach,
\begin{quote} It is not (strictly speaking) true that we observe particles.
Rather, there are `observation events', and these observation events
are consistent (to a good degree of accuracy) with the supposition that
they are brought about by (localizable)
particles.\cite{halvorsonclifton}
\end{quote}
From the perspective of this paper, the problem with this approach is
its a priori assumption that what we measure are always exactly
localised operators.  This is, of course, an interpretive axiom of AQFT
as it is often presented, but it effectively assumes the presence of
outside observers whose measurements cannot be treated within the
ordinary dynamics of the QFT.  We shall instead construct an account
which treats observers as part of the internal dynamics of the system
(although, apart from that difference of emphasis, the solution below
will be rather similar in character to that of Halvorson and Clifton).

If we wish instead to treat our QFT as a
closed system, and the measurement process as part of the internal
dynamics of that system, then it is an open question whether or not we must treat 
question our measurements as localised.  Furthermore, we have at least some
reason to think that the answer to the question is negative --- for we
believe our measuring devices (including ourselves) to be made out of
particles, and we have already noted the fact that particles are never
exactly localised.  

But does it even make sense to consider non-localized measuring devices?
To see that it does (at least in some sense), we return (again) to non-relativistic quantum
mechanics.  In NRQM it is unproblematic, at least in principle, to
construct measurement devices out of atoms.  Such devices usually have
rather large masses, so despite the uncertainty principle they can be
put in states whose position and momentum are both very sharp.  In
particular, if a measurement device's momentum is sharply peaked around
zero then the system will effectively remain in a single well-determined
region.

But will it exactly remain in that region?  The mathematics says not,
for it is mathematically impossible for any centre-of-mass wavefunction
to remain localised in a finite spatial region for any finite period of
time.  (A rather general proof of this result has been given by
Hegerfeldt \citeyear{hegerfeldta,hegerfeldtb}, and is discussed by \citeN{halvorsonclifton}.)  This
is implicit, in fact, in the wave-functions we usually use in NRQM to describe
`localised' systems: they are generally described by fairly sharp
Gaussians, but any Gaussian --- no matter how sharp --- is non-zero
everywhere in space.

Does this mean that the Geiger counter in my lab has finite amplitude to
be in Andromeda?  No, for the representation of a quantum object by a
Gaussian state relies on the fact that the object is not entangled with
its environment.  But the part of the wavefunction representing the
device as (for instance) in the walls of the lab will obviously
interact with its environment in a rather different way from the part
which represents the device as in the middle of the lab.  As such, the
device should not strictly speaking be represented by a pure state at
all, but is instead entangled with its surroundings.

The reason why we do not in practice need to allow for this entanglement
is that it is ludicrously small.  Consider, for instance, a 
device of mass $m$ whose centre-of-mass wave-function has Gaussian form
\be \psi \propto \exp(-r^2/2 L^2).\ee
If the device is allowed to evolve freely, after some time $t$ the
centre-of-mass wave-function will evolve to 
\be \psi(r,t) \propto \exp(-r^2/2 L_t^2)\ee
where $L_t^2=L^2+it/2 m$.  
The probability density for finding the particle a distance $r$ from the
origin at time $t$ is
\be |\psi(r,t)|^2 \propto \exp (-r^2/D_t^2)\ee
where $D_t^2=(L^4+t^2/4 m^2)/L^2$; hence the device's wave-function will
have essentially constant half-width as long as $t\ll 2m L^2$.  For a
one-kilogram device whose initial half-width is $10^{-11} \mathrm{m}$,
this means that the half-width will stay approximately constant on
timescales of order $10^6$ years.  The probability of finding the
particle in a given volume a distance $r$ from the origin, compared
with the probability of finding it in a similar-sized volume around the
origin, is $|\psi(r,0)|^2/|\psi(0,0)|^2=\exp(-r^2/L^2).$  If we take $r=1
\mu \mathrm{m}$, then this probability ratio is approximately equal to  
$10^{-10^{10}}$.  Only if we are working to $10^{10}$-significant-figure accuracy do we
need to allow for terms like this, and calculations are seldom done to
such accuracy!

The truth, of course, is that the real state of any realistic model of
NRQM is an entangled mess.  It contains \emph{no} exactly localised
objects and \emph{no} truly isolated subsystems, and there is \emph{no}
chance of working out its exact dynamics.  Our actual strategy is to find a decomposition into 
subsystems such that they are \emph{approximately} isolated, and to
accept the errors that may result from this strategy.

We can quantify this by means of the Hilbert-space norm.  Let \ket{\psi}
be the actual state of the system, and $\op{U}(t)$ the unitary operator
generating its actual time evolution.  Then to say that the state is
approximately two isolated subsystems (during a period of length T) is to say that:
\begin{enumerate}
\item There is some state $\ket{\sim \psi}=\tpk{\psi_1}{\psi_2}$ such
that $|\bk{\psi}{\sim\psi}|\simeq1$;
\item There is some unitary operator
$\op{U}_\sim(t)=\op{U}_1(t)\otimes\op{U}_2(t)$, such that, for $t<T$,
$|\matel{\psi}{\opad{U}(t)\op{U}_\sim(t)}{\sim\psi}|\simeq
1$.\footnote{There are two simplifications in this account.  Firstly, in
general a state of NRQM contains large numbers of identical particles,
so the tensor product in (1) should be symmetrised and/or
antisymmetrised to allow for these particles.  Secondly, many subsystems
(particularly those of macroscopic dimension) undergo decoherence, and
hence cannot literally be said to be isolated in the sense used here.  For systems whose
dynamics are regular (\ie non-chaotic) this is not particularly significant: decoherence will
be significant if the system is not in some state of a certain preferred
basis, but if it is in such a state then the isolated quantum dynamics
will fairly accurately predict its evolution.  For chaotic systems we cannot ignore decoherence
in this manner (see
\citeN{zurek98} for details).}
\end{enumerate}
(Note that we are not requiring $\op{U}_\sim(t)\simeq \op{U}(t)$ in any
sense; our discussion is entirely state-dependent.)

If our standard description of a certain complex object (a DNA strand, a Geiger
counter, a computer, whatever) describes it as an isolated object, and
if isolated subsystems are only ever approximations to the `true'
quantum state, it might seem that we should say something like `complex objects 
don't exist, they only approximately exist'.  But this is to
misunderstand the status of complex objects in physics.  Such objects
are actually identified by their structural properties (interacting with
other DNA strands in a certain way, detecting radiation, calculating
$\pi$, whatever) and such structural properties can tolerate small
amounts of noise.  (Put another way, an object which operates like a
computer with $99.999999\ldots\%$ accuracy is still a computer!)  From this
viewpoint, to say that a certain complex object is present is to say
that making that approximation is an extremely effective method of
analysing the dynamics.  (See \citeN{wallacestructure} for a more
extended discussion of this approach to complex objects).

Localisation enters this framework as a pragmatic criterion for
isolation.  If, in the state $\ket{\sim\psi}=\tpk{\psi_1}{\psi_2}$, the
states \ket{\psi_1} and \ket{\psi_2} have negligible amplitude to be
in the same spatial location, and if the actual dynamics (as described
by $\op{U}(t)$) are generated by a Hamiltonian in which the 
inter-particle forces drop off strongly with distance, then we can apply
the isolated-subsystem approximation with considerable confidence as to
its accuracy; conversely, if at some point this approximation predicts
that $\ket{\psi_1}$ and $\ket{\psi_2}$ have come to be localised, with
high amplitude, in the vicinity of one another, then we should abandon
the approximation from that point onwards.

This means that --- not withstanding the existence in NRQM of a
precisely definable position operator --- we should treat the spatial
location of a given object as approximate.  It is not that my Geiger
counter is localised in a number of exact locations with different
amplitudes for each; rather, it is determinately\footnote{No comment on the measurement
problem is intended here; Everettians should read this as `determinately localised relative
to the branch which we are considering'.} localised in some approximately-defined location.  The
exponentially small tails of the centre-of-mass wavefunction are totally irrelevant to any considerations
of the counter's spatial location, for it is only in the approximation where these are neglected that
we can treat the Geiger counter as isolated anyway.

Before applying these ideas to QFT, we should acknowledge a lacuna in
the argument: we have assumed that the Hilbert-space norm is the
appropriate measure of closeness of approximation, but what justifies
this assumption? 
From a purely mathematical point of view, to be sure, it is an extremely obvious choice:
it is a very natural metric on the space of states, and the unitarity of Hilbert-space dynamics
means that the Hilbert-space distance between two points is time-invariant.  But the situation
is more complicated than this: \citeN{reutsche} (writing in the context of the modal interpretation
of quantum mechanics) points out that whilst a state like
\be \alpha \ket{\mathrm{Spin\, up}}+\beta \ket{\mathrm{Spin\, down}}\ee
(where $\alpha\simeq 1$ and $\beta \simeq 0$) may legitimately be regarded as `almost' a spin-up
state, it is highly problematic to suppose that a macroscopic superposition like
\be \alpha \ket{\mathrm{Living\, cat}}+\beta \ket{\mathrm{Dead\, cat}}\ee
is almost a live cat, whatever the values of $\alpha$ and $\beta$.
She describes this sort of approximation (borrowing her terminology from \citeNP{teller}) as 
``ontological distortion'', as against the mere ``numerical distortion'' that occurs when, say, 
the position of a classical particle is predicted with slight
inaccuracy: a macroscopic superposition is \emph{not} (prima facie) even approximately the
same as a system in a macroscopically definite state, 
but an ontologically new and
problematic entity.  

It is now clear that justifying the use of the Hilbert-space norm in
this way takes us into the murky waters of the quantum measurement
problem.  We can see this another way by 
considering how easily an instrumentalist could dismiss our
worries: from his perspective, the empirical content of the Hilbert-space norm is simply given by
the probabilities of measurements to yield certain results, and states differing only very slightly
in Hilbert space will give only slightly different probabilities for a given outcome on measurement.  
It is the illegitimacy of using ``measurement'' as a primitive concept
in our approach which causes the difficulties.

In fact (staying in the non-relativistic domain) our problem is very closely related to 
some standard foundational questions in a number of interpretations:
\begin{itemize}
\item In the modal interpretation \cite{vanfraassen,dieksvermaas} the question is essentially 
equivalent to the
Albert-Loewer (Albert and Loewer \citeyearNP{albertloewer1990,albertloewer1991}; 
\citeNP{albert1992}) problem of vagueness, discussed by \citeN{reutsche} 
(see also references therein):
the idea of a modal interpretation is to pick out a certain set of properties as definite, whilst
it seems likely that the properties actually picked out are slightly different from the 
``right'' ones --- where `slightly' means `slightly in Hilbert-space
norm'.  The standard (`vague property') response to this is to replace precise properties
(like spatial localisation) with a family of `vague' properties, all
very close in Hilbert-space norm to the `precise' property under
consideration.\footnote{In fact, the considerations of this section
suggest a more serious problem for the modal interpretation: the very
concept of macroscopic (non-relativistic) subsystems is threatened by
wave-function spreading.  \citeN{reutsche} discusses a library book,
localised approximately in her office but with exponential tails giving
it a nonzero amplitude to be anywhere in space, and claims that such a
book `is in no way compatible with my ordinary notion of \emph{located
in my office}' (\cite[p.\,233]{reutsche}; italics hers) --- but in fact,
the interaction of those tails with the walls of the office will lead to
the sort of rapid interaction that removes any prospect of
distinguishing `book' and `wall' subsystems.}
\item In collapse interpretations such as those of \citeN{GRW} and
\citeN{pearle} the question is equivalent to the `problem of tails'
\cite{albertloewer1996}: the collapse mechanism is generally chosen to
cause the wave-function to be sharply peaked around a given spatial
position but it is impossible to construct mechanisms which prevent the
wave-function having exponential tails extending to infinity.  The
problem of tails can be addressed \cite{albertloewer1996} by postulating 
some rule such as ``a particle is localised in spatial region \mc{R} if the squared
modulus of the wave-function integrated over \mc{R} exceeds $(1-p)$'', 
where $p$ is taken as small.  There has been recent controversy
(\citeNP{lewis,ghirardi98,bassi99}, Clifton and 
Morton \citeyearNP{cliftonmonton1,cliftonmonton2}) as to 
whether this rule should be understood as
a metaphysical principle (which leads to difficulties both because of
the arbitrariness of $p$ and because of alleged failures of the
arithmetic of macroscopic objects) or, as advocated by \citeN{cliftonmonton1}, as a 
mere \textit{facon de parler} useful in describing the wave-function (in
which case there is presumably need for a defence of treating the tails
as irrelevant to observational predictions).
\item From the viewpoint of the Everett interpretation (at least in the form advocated by
the author \cite{wallacestructure,wallaceworlds}, which is essentially
that developed by Zurek(most explicitly in \citeNP{zurekprob}) and 
Saunders \citeyear{saundersdecoherence,saunderstense,saundersprob}) our
problem is that of justifying the neglect of parts of the wave-function
of extremely low weight.  This is essentially the probability problem of
Everett interpretations: what justifies our strategy of disregarding
low-weight branches when making decisions?
\end{itemize}
Although expanding on these rather brief comments would take us deep
into interpretational questions and far beyond the scope of this paper,
it seems likely that (at least for the interpretations above) a
resolution of the quantum measurement problem would include a
justification of the Hilbert-space norm as the appropriate measure of approximation.
This seems particularly certain in the Everett interpretation, which is probably 
the interpretational program most naturally extendible to QFT and most
in keeping with the spirit of this paper.

Given an interpretational justification of our use of the Hilbert-space
norm, the concepts above carry through to QFT with only minor changes.  The 
dynamical impossibility of constructing exactly isolated subsystems in
NRQM
becomes a kinematic impossibility in QFT if we consider the spatially
localised subsystems $\mc{H}_{\Sigma_i}$, but for essentially the same
reasons: interactions between those subsystems are so prevalent that the
energy cost of constructing non-entangled states would take us out of
the domain of validity of the QFT altogether.  Also, in QFT it is even
more apparent than in NRQM that objects can only be defined
approximately unless they are isolated, for all that exist in QFT are
the field observables, excitations of them, and patterns in those
excitations, and if two patterns overlap and interact then there can be
no exact criteria for individuating them.  And in QFT there are no
action-at-a-distance interactions, so spatial location is an even better
criterion for isolation: two subsystems can be treated as isolated iff
they are effectively localised in spatially separated regions.

Consequently, the approximate criteria developed above for NRQM apply
just as well in QFT.  In particular, QFT particles (unlike NRQM
particles, which are part of the basic formalism) are themselves just
certain patterns of excitations in the fields.  As such, the appropriate
concept of localisation for them is approximate, and so perfectly
adequately treated by the methods of section \ref{effloc}.  If we say,
for instance, that a particle is localised in (spatial) region $\Sigma_i$,
and that the region in the vicinity of $\Sigma_i$ is empty, then what
we mean is:
\begin{itemize}
\item that the field state restricted to $\Sigma_i$ coincides almost exactly
with the restriction to $\Sigma_i$ of some state in the one-particle Hilbert
space $\mc{H}_{1P}$;
\item that in the region surrounding $\Sigma_i$ the field state almost
coincides with the vacuum.
\end{itemize}
These requirements are perfectly consistent with the actual field state
being that of a particle (or of a particle together with some other
excitations, created by operators strictly localised far from
$\Sigma_i$); they are also consistent with the actual state itself being
strictly (\iec, Knight-) localised in $\Sigma_i$.  The point is, we neither know nor
care: different ways of realising the requirements just mean different 
--- but tiny --- amounts of noise added to our highly-accurate
description of the quantum state and of its dynamics.  Ditto for
particle \emph{detectors}: there is no more need (and no more prospect)
to regard them as exactly localised, provided that they are very well
localised.  If we model such a detector as made out of particles then
that model will describe it as differing from the vacuum all the way to
infinity, but it will also describe those differences as incredibly
small beyond a certain very-well-defined, but not \emph{precisely}
defined, region.  Then if the actual state of the QFT is almost the
same as that of the model within that region, and is almost the same as the vacuum outside that
region, then the criterion has been met
for us to say that the region contains a detector.\footnote{We could, to be sure, define a detector as being present if the QFT
state is almost the same as the model state within the localisation
region, and \emph{exactly} the same as the vacuum outside it --- but
what would be the justification for so doing?  We don't have any reason
to suppose that actual lab detectors are precisely localised in this
way.}

\section{Conclusion}\label{conclusion}

This completes our analysis of Lagrangian QFTs.  We have argued that
such QFTs can be made into perfectly well-defined quantum theories
provided we take the high-energy cutoff absolutely seriously; that the
multiple ways of doing this are not in conflict provided that we
understand them as approximations to the structure of some deeper, as
yet unknown theory; that the existence of inequivalent representations
is not a problem; that a concept of localisation can be defined for such
theories which is adequate to analyse at least some of the practical
problems with which we are confronted; and that the inexactness inherent
in that concept is neither unique to relativistic quantum mechanics, nor
in any way problematic.

If there is one underlying theme to the approach to QFT advocated in
this paper, it is this: the sort of information which we are interested in getting from physical
theories is structural information.  The reason that states with isolated subsystems
approximate real quantum states well (both kinematically and dynamically) is because 
the two have virtually the same structure; the reason that we can be
indifferent as to which concrete realisation of a scale-$l$ AQFT to use
is that structurally they are all virtually identical (at least until we
get to scales of order $l$, at which point we no longer believe that any
of them give us valid structural information).

The need to understand theories in structural terms is not restricted to
QFT, but it is more starkly obvious there.  It is not just that QFT is
strictly speaking false, or that it is expected to be replaced someday
by a deeper `theory X'; that is true for virtually every physical theory we
study.  It is rather that ---
if Wilson's approach to
renormalization is taken seriously --- QFT only makes sense if we
include in it some vestigial aspects of the very theory which we expect
to replace it.  

From this viewpoint, we can see that Lagrangian QFT (as I have defended
it) is not really in conflict with AQFT at all.  Success in the AQFT
program would leave us with a field theory exactly defined on all
scales, and such a theory would be a perfectly valid choice for `theory
X': furthermore, even if we found such an exact QFT it would not prevent us from 
defining low-energy, `effective' QFTs --- which would not be well defined without a cutoff; nor,
probably, would it obviate the need for these theories in describing
certain low-energy limits of X.  

If AQFT has any rival programs, in fact, 
they are string theory and other theory-of-everything
candidates.  Success in any of these programs would, of course,
revolutionise physics, but that success would scarcely change the
current status of Lagrangian QFT: as an inherently approximate, but
nonetheless extraordinarily powerful tool to analyse the deep structure
of the world.

\subsection*{Acknowledgements}
I would like to thank Simon Saunders, whose own work on the foundations of QFT
prompted my interest in the subject; and Jeremy Butterfield for many useful comments
on an earlier draft of this paper.  I have also benefitted greatly from conversations with
Ian Aitchison, James Binney, Katherine Brading, Keith Burnett, Peter Morgan, and Andrew 
Steane.

\end{document}